\documentclass[pra,aps,groupaddress,showpacs,twocolumn]{revtex4-1}
\usepackage{epsfig,amsmath,graphicx,amssymb,overpic}
\def\be{\begin{equation}}
\def\ee{\end{equation}}
\def\bee{\begin{eqnarray}}
\def\ene{\end{eqnarray}}
\def\bes{\begin{subequations}}
\def\ees{\end{subequations}}

\newcommand{\br}{{\bf r}}
\newcommand{\bv}{{\bf v}}

\newcommand{\PT}{\mathcal{PT}}

\begin{document}
\title{On stable solitons and interactions of the generalized Gross-Pitaevskii \\ equation with  $\PT$- and non-$\PT$-symmetric potentials}
\author{Zhenya Yan}
\email{Author to whom correspondence should be addressed. Electronic mail: zyyan@mmrc.iss.ac.cn.}
\author{Yong Chen}
\author{Zichao Wen}
\affiliation{Key Laboratory of Mathematics Mechanization, Institute of Systems
Science, AMSS, Chinese Academy of Sciences, Beijing 100190, China \\
School of Mathematical Sciences, University of Chinese Academy of Sciences, Beijing 100049, China \vspace{0.1in}}

\date{\vspace{0.1in} 29 March 2016, CHAOS {\bf 26}, 083109 (2016)}


\begin{abstract}

We report the bright solitons of the generalized Gross-Pitaevskii  (GP) equation with some types of physically relevant parity-time- ($\PT$-) and non-$\PT$-symmetric potentials. We find that the constant momentum coefficient $\Gamma$ can modulate the linear stability and complicated transverse power-flows (not always from the gain toward loss) of nonlinear modes. However, the varying momentum coefficient $\Gamma(x)$ can modulate both unbroken linear $\PT$-symmetric phases and stability of nonlinear modes. Particularly, the nonlinearity can excite the unstable linear mode (i.e., broken linear $\PT$-symmetric phase) to stable nonlinear modes. Moreover, we also find stable bright solitons in the presence of non-$\PT$-symmetric harmonic-Gaussian potential. The interactions of two bright solitons are also illustrated in $\PT$-symmetric potentials. Finally, we consider nonlinear modes and transverse power-flows in the three-dimensional (3D) GP equation with the generalized $\PT$-symmetric Scarff-II potential.
\end{abstract}

\maketitle



\textbf{The study of stable solitons is of the important significance in many fields of nonlinear science such as nonlinear optics, Bose-Einstein condensates, fluid mechanics, plasmas physics, ocean, etc. The gain-and/or-loss distribution can usually generate to unstable solitons. In 1998, Bender {\it et al.} first presented the non-Hermitian parity-time- ($\PT$-) symmetric Hamiltonians possessing entirely real spectra for some parameters, which differs from usual Hermitian Hamiltonians. In fact, the imaginary parts of $\PT$-symmetric potentials can be regarded as the gain-and-loss distribution in linear eigenvalue problems. After that, the $\PT$-symmetric potentials
were introduced in the nonlinear Schr\"odinger equation such that the stable solitons could also be found since the imaginary parts of $\PT$-symmetric potentials can always be balanced. A nature problem is whether some stable solitons may also exist in other nonlinear wave models with $\PT$-symmetirc potentials. In this paper, we report bright solitons of the
generalized GP equation with physically relevant $\PT$- and non-$\PT$-symmetric potentials. We find that the constant momentum coefficient can modulate the linear stability and complicated transverse power-flows (not always from the gain toward loss) of nonlinear modes. However, the varying momentum coefficient $\Gamma(x)$ can modulate both unbroken linear $\PT$-symmetric phases and stability of nonlinear modes. Particularly, the nonlinearity can excite the unstable linear mode (i.e., broken linear $\PT$-symmetric phase) to stable nonlinear modes. Moreover, we also find the stable bright solitons in the presence of non-$\PT$-symmetric harmonic-Gaussian potential. Finally, we consider nonlinear modes and their transverse power-flows in three-dimensional GP equation with the generalized $\PT$-symmetric Scarff-II potential. These results may excite the  potential experiments and applications in Bose-Einstein condensates, nonlinear optics, and other relevant fields.}

\section{Introduction}

The theory of classical quantum mechanics usually requires that the observable Hamiltonian is Hermitian, which must admit entirely real energy spectra~\cite{qm}. The Dirac Hermiticity is, in fact, a sufficient but not
necessary condition to guarantee that its all energy spectra are real since Bender and Boettcher~\cite{Bender98} first showed
that the non-Hermitian Hamiltonian, $H=p^2+x^{2}(ix)^{\epsilon}$\, ($\epsilon$ is real) still possessed entirely real eigenvalues for $\epsilon\geq 0$. The novel results are mainly due to the introduction of parity ($\mathcal{P}$)-time ($\mathcal{T}$) reflection symmetry, where two operators are defined as~\cite{Bender2}: the linear parity operator $\mathcal{P}: x\rightarrow -x$ and the antilinear time-reversal operator $\mathcal{T}: i\rightarrow -i$. In other words, the $\PT$ symmetricity may be regarded as a `good property' easily used to seek for the parameter domains of unbroken $\PT$-symmetric phase for the non-Hermitian Hamiltonians. In terms of this $\PT$-symmetric principle, the one-dimensional linear Schr\"odinger operator $\hat H=-\partial_x^2+U(x)$ is $\PT$-symmetric provided that $U(x)=U^{*}(-x)$, that is, $U(x)$ has the even real part and the odd imaginary part for space~\cite{Bender2}, where $U(x)$ is also called the refractive-index in optical fibre.

Recently, the various complex $\PT$-symmetric phenomena have been observed in some fields~\cite{Exp1, Exp2, exp2a, exp5, Exp3,Exp4}. Moreover, some stable nonlinear modes were excited in various $\PT$-symmetric potentials such as the Scarff-II potential~\cite{ptsf, yanpre15}, optical lattice potential~\cite{ptsf2}, harmonic-Gaussian potential~\cite{harm1},  harmonic potential~\cite{harm-l},  Gaussian potential~\cite{harm1,gau2}, sextic anharmonic double-well potential~\cite{anharm}, time-dependent harmonic-Gaussian potential~\cite{yanpra15}, the double-delta potential~\cite{delta}, super-Gaussian potential~\cite{sg}, and etc.~\cite{other}. Most of linear Hamiltonian operators with one particle in these systems are of the same form $\hat H$, i.e., the sum of kinetic and $\PT$-symmetric potential energy operators~\cite{Bender2}. More recently, the
stable solitons of the three-order NLS equation with $\PT$-symmetric potentials have been investigated~\cite{yan16}.

 If one considers one more term (momentum operator) $i\Gamma(x)\partial_x$ with the coefficient $\Gamma(x)$ being a real-valued parameter or function of space, then the generalized Hamiltonian operator is generated as  $\mathcal{H}=-\partial_x^2+i\Gamma(x)\partial_x+U(x)$. Recently, the generalized Hamiltonian operators with various real-valued potentials $U(x)$ were introduced in the nonlinear Schr\"odinger (or Gross-Pitaevskii) equation to excite stable wave propagations of nonlinear modes, such as the periodic~\cite{rbec, rbecp1, rbecp2}, anharmonic~\cite{rbec}, and two narrow Gaussian barrier~\cite{rbecb} potentials. However, to the best of our knowledge, the above-mentioned operator $\mathcal{H}$ with complex potentials (e.g., $\PT$-symmetric potentials) has not been introduced into nonlinear physical models. Thus, the first goal of the present work is to replace real potentials~\cite{rbec, rbecp1, rbecp2,rbecb} with some physically relevant $\PT$-symmetric potentials (e.g., Scarff-II, generalized Scarff-II double-well, and harmonic-Gaussian potentials) into the generalized NLS/GP equation~\cite{rbec}. Moreover, we find that the momentum coefficient and gain-and-loss distribution have the strong effects on nonlinear modes and transverse power-flows. If the momentum coefficient is dependent on space, then another aim of this paper is to study the effects of the varying momentum coefficient $\Gamma(x)$ and $\PT$-symmetric potentials on linear and nonlinear modes as well as transverse power-flows such that some interesting results are found.

The rest of this paper is arranged as follows. In Sec. II we introduce the generalized GP (NLS) equation with the $\PT$-symmetric potentials and give the general theory. In Sec. III, we
consider some physically interesting potentials, in which
we study the bright solitons for both self-focusing (attractive) and defocusing (repulsive) cases and their stability in the domains of both unbroken and broken $\PT$-symmetric phases. Moreover, we find that the nonlinearity can excite stable nonlinear modes in the domain of broken $\PT$-symmetric phase. The momentum coefficient strongly modulates the transverse power-flows such that the transverse power flowing directions are complicated. Moreover, we find the stable solitons in the presence of non-$\PT$-symmetric harmonic-Gaussian potential. In Sec. IV, we investigate the effect of varying  momentum coefficient on the spectra of the linear problem and stability of nonlinear modes. In Sec. V, we consider the 3D nonlinear model with the generalized $\PT$-symmetric Scarff-II potential. Finally, some results are given in Sec. VI.

\section{Nonlinear model with $\PT$-symmetric potentials}

\subsection{The nonlinear  model}

We here focus on the one-dimensional generalized Gross-Pitaevskii (or NLS) equation with the $\PT$-symmetric potentials as
\bee\label{nls}
 i\partial_t\psi\!=\!\left(\!-\frac12\partial_x^2+i\Gamma(x) \partial_x+V(x)+iW(x)-g|\psi|^2\right)\!\psi,\quad
\ene
where $\partial_t=\partial/\partial t$, $\psi\equiv \psi(x, t)$ is a complex condensate wave function of $x,t$,
the term $i\Gamma(x)\partial_x$ is called the momentum operator, the
$\PT$-symmetric potential requires that $V(x)=V(-x)$ and $W(x)=-W(-x)$ describing the real-valued potential and gain-loss distribution, respectively, and $g>0$\, (or $<0)$ is real-valued attractive (or repulsive) nonlinear interactions, respectively. Eq.~(\ref{nls}) can be regarded as a special case of the coupled GP equations with
$\PT$-symmetric potentials describing the quasi-one-dimensional spin-orbit Bose-Einstein condensate by the spinor $\Psi=(\psi_1, \, \psi_2)^T$~\cite{so1,so2,so3}
\bee\label{nls2}
 i\partial_t\Psi=\frac{1}{2}\left(\frac{1}{i}\partial_x-\kappa(x)\sigma_1\right)^2\!\Psi+\frac{\Omega}{2} \sigma_3\Psi-(\Psi^{\dag}\Psi)\Psi,\qquad
\ene
with $\sigma_{1,3}$ being Pauli matrices, that is, if we set $\psi=\psi_1=\psi_2$,\, zero-Zeeman splitting $\Omega=0$, \, $2\Gamma(x)=\kappa(x)$, add the $\PT$-symmetric potentials
$V(x)+iW(x)-\frac12[i\kappa_x+\kappa^2(x)]$ and change the nonlinearity $1$ into a constant $g$ in system (\ref{nls2}), then system (\ref{nls2}) reduces to the single model (\ref{nls}) that we will consider. Thus for the possible experimental implementation of the studied  model (\ref{nls}), one may design it in terms of the similar results (see, e.g., Refs.~\cite{so1,so2,so3}).

It is easy to show that Eq.~(\ref{nls}) is invariant under the $\PT$-symmetric transformation if the complex potential $[V(x)+iW(x)]$ is $\PT$-symmetric and $\Gamma(x)$ is an even function of space, where ${\mathcal P}$ and ${\mathcal T}$ operators are defined by ${\mathcal P}:\, x\to -x$ and ${\mathcal T}: t\to-t,\,  i\to-i$.
Equation (\ref{nls}) is associated with a variational principle $i\psi_t=\delta \mathcal{H}(\psi)/(\delta \psi^{*})$ with the Hamiltonian $\mathcal{H}(\psi)=\int_{-\infty}^{+\infty}\{\frac12|\psi_x|^2
 +i\frac{\Gamma(x)}{2}\left(\psi^{*}\psi_x-\psi\psi^{*}_x\right)
+[V(x)+iW(x)]|\psi|^2-\frac{g}{2}|\psi|^4\}dx$,
where the asterisk stands for the complex conjugate. The quasi-power and power of Eq.~(\ref{nls}) are given by $Q(t)=\int_{-\infty}^{+\infty}\psi(x,t)\psi^{*}(-x,t)dx$ and $P(t)=\int_{-\infty}^{+\infty}|\psi(x,t)|^2dx$, respectively. One can readily know that $Q_t=i\int_{-\infty}^{+\infty}g\psi(x,t)\psi^{*}(-x,t)[|\psi(x,t)|^2-|\psi(-x,t)|^2]dx$ and $P_t=
2\int_{-\infty}^{+\infty}W(x)|\psi(x,t)|^2dx$. In the absence of gain-and-loss distribution $W(x)$, Eq.~(\ref{nls}) has been studied in the presence of constant $\Gamma(x)$ and the absence of gain-and-loss distribution $W(x)$~\cite{rbec, rbecp1, rbecp2, rbecb}.

\subsection{General theory for the stationary problem}

We are now interested in the localized stationary solution of Eq.~(\ref{nls}) as $\psi(x,t)=\phi(x)e^{i\mu t}$, where $\mu$ denotes the real chemical potential and the complex wave function $\phi(x)$ ($\lim_{|x|\to\infty} \phi(x)=0$) satisfies the generalized stationary GP equation
\bee\label{ode}
\left(\!-\frac12\frac{d^2}{dx^2}\!+\!i\Gamma(x)\frac{d}{dx}\!+\!V(x)\!+\!iW(x)\!-\!g|\phi|^2\!+\!\mu\!\right)\!\phi\!=\!0.\,\,
\ene
 Eq.~(\ref{ode}) with $g=0$ becomes the linear eigenvalue problem. In the presence of nonlinearity, there exist two cases for the study of solutions of Eq.~(\ref{ode}): i) if $\phi(x)$ is a real-valued function, then we have the solution of Eq.~(\ref{ode})
\bee \label{solua}
 \phi(x)=c\exp\left[-\int_0^xW(s)/\Gamma(s) ds\right],
\ene
with $c\not=0$ and the condition linking the potential and gain-or-loss distribution being
\bee
 V(x)\!=\!\frac{W^2\!-\!\Gamma W_x\!+\!W \Gamma_x}{2\Gamma^2}\!+\!gc^2\!\exp\!\left[\!-\!2\!\int_0^x\!\!\frac{W(s)}{\Gamma(s)}ds\right]\!-\!\mu,\,\,\,\,\,\,\,\,\,\,
\ene
ii) if the function $\phi(x)$ is complex, that is
\bee \label{solub}
 \phi(x)=\hat\phi(x)\exp\left[i\int^x_0v(s)ds\right],
\ene
where the real function $\hat\phi(x)$ is the amplitude, and the real function $v(x)$ is the hydrodynamic velocity, then we substitute Eq.~(\ref{solub}) into Eq.~(\ref{ode}) to yield the relation linking the hydrodynamic velocity
\bee\label{ode1}
 v(x)=2\hat\phi^{-2}(x)\int^x_0[W(s)\hat\phi^2(s)+\Gamma(s)(\hat\phi^2(s))_s]ds,
\ene
and the amplitude satisfying
\bee \label{ode2}
 \frac{\hat\phi_{xx}(x)}{\hat\phi(x)}\!+\!2g\hat\phi^2(x)\!-\!(v(x)\!-\!\Gamma(x))^2\!-\!2V(x)\!=\!2\mu\!-\!\Gamma^2(x).\,\,\,\,\,\,\,
\ene

 For the given $\PT$-symmetric potentials, one may solve Eq.~(\ref{ode}) (or Eqs.~(\ref{ode1}) and (\ref{ode2})) by analytical or numerical methods. Hence we have found the nonlinear modes in the stationary  form $\psi(x,t)=\phi(x)e^{i\mu t}$ with $\phi(x)$ given by Eq.~(\ref{solub}). To further study the linear stability of the above-obtained nonlinear stationary solutions, we considered a perturbed solution~\cite{stable, yang}
\begin{equation}
\label{pert}
\psi(x,t)=\left\{\phi(x)+\epsilon \left[F(x)e^{i\delta t}\!+\!G^*(x)e^{-i\delta^* t}\right]\right\}e^{i\mu t},
\end{equation}
where $\phi(x)e^{i\mu t}$ is a stationary solution of Eq.~(\ref{nls}), $\epsilon\ll 1$, and $F(x)$ and $G(x)$ are the eigenfunctions of the linearized eigenvalue problem. Substituting Eq.~(\ref{pert}) into Eq.~(\ref{nls}) and linearizing with respect to $\epsilon$, we obtain the following linear eigenvalue problem
\begin{eqnarray} \label{st}
\left(\begin{array}{cc}   L & g\phi^2(x) \vspace{0.05in}\\   -g\phi^{*2}(x) & -L^* \\  \end{array}\right)
\left(  \begin{array}{c}    F(x) \vspace{0.05in} \\    G(x) \\  \end{array} \right)
=\delta \left(  \begin{array}{c}   F(x) \vspace{0.05in}\\    G(x) \\  \end{array}\right),
\label{stable}
\end{eqnarray}
where $L=\frac12\partial^2_x-i\Gamma(x)\partial_x-[V(x)+iW(x)]+2g|\phi(x)|^2-\mu$. It is easy to see that
the $\PT$-symmetric nonlinear modes are linearly stable if $\delta$ has no imaginary component, otherwise they are linearly
unstable.

\section{$\PT$-symmetric linear and nonlinear modes}

In this section we mainly consider the simple case $\Gamma(x)=\Gamma={\rm const.}$ in Eq.~(\ref{nls}).

\subsection{$\PT$-symmetric Scarff-II potential}

We firstly consider the physically interesting potential in Eq.~(\ref{nls}) as the $\PT$-symmetric Scarff-II potential~\cite{scarff}
\bee\label{ps}
 V(x)=V_0\,{\rm sech}^2x, \quad  W(x)=W_0\,{\rm sech}x\tanh x,
\ene
where the real constants $V_0<0$ and $W_0$ can modulate  amplitudes of the reflectionless potential $V(x)$~\cite{rp} and gain-and-loss distribution $W(x)$, respectively. These two functions are bounded  and $V(x), W(x)\to 0$ as $|x|\to \infty$.   Moreover, the gain-and-loss distribution can be always balanced in Eq.~(\ref{nls}) since $\int^{+\infty}_{-\infty}W(x)dx=0$.

\subsubsection{Broken/unbroken linear $\PT$-symmetric phases}

 \begin{figure}[!t]
 	\begin{center}
 	\vspace{0.05in}
 	\hspace{-0.05in}{\scalebox{0.24}[0.24]{\includegraphics{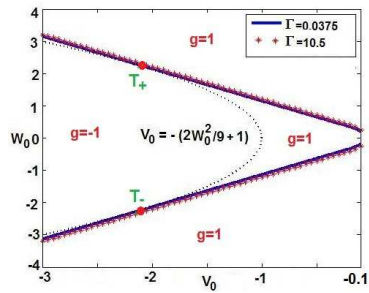}}}
 	\end{center}
 	\vspace{-0.15in} \caption{\small (color online). The profile of phase transitions for the linear operator $L_s$ (\ref{ls}) with (\ref{ps}). The unbroken (broken) $\PT$-symmetric phase is in the domain inside (outside) two symmetric phase breaking lines for every different frequency $\Gamma=0.0375, 10.5$. The domain of existence condition for bright solitons (\ref{solu}) for the attractive $g=1$ (repulsive $g=-1$) cases is outside (inside) the parabola $V_0=-(2W_0^2/9+1)$. The tangent points are $T_{\pm}=(-2.125,\, \pm 2.25)$. }
 	\label{fig-spectra1}
 \end{figure}

We firstly consider the linear eigenvalue problem with $\PT$-symmetric Scarff-II potential (\ref{ps}) in Eq.~(\ref{nls}) as
\bee \label{ls}
 L_s\Phi(x)=\lambda\Phi(x),\, L_s=-\frac{1}{2}\partial_x^2\!+\!i\Gamma\partial_x\!+\! V(x)\!+\! iW(x),\qquad
\ene
where $\lambda$ and $\Phi(x)$ are the eigenvalue and engenfunction, respectively, and $\lim_{|x|\to \infty}\Phi(x)=0$.  In the absence of momentum term  $\Gamma=0$, we know that the linear problem (\ref{ls}) reduces to the usual Hamiltonian operator with $\PT$-symmetric Scarff-II potential (\ref{ps}):
$L_0\Phi_0(x)=\lambda_0\Phi_0(x), \, L_0=-\frac{1}{2}\partial_x^2\!+\! V(x)\!+\! iW(x)$,
where $\lambda_0$ and $\Phi_0(x)$ are the eigenvalue and eigenfunction, which can be shown to admit entirely real (discrete) spectra provided that the parameters $V_0<0$ and $W_0$ satisfy~\cite{scarff}
 \bee \label{ptc}
|W_0|\leq \frac{1}{8}-V_0.
\ene
In fact, Eq.~(\ref{ls}) admits the exact state solution $\Phi(x)=c\,{\rm sech}(x)\exp[-2iW_0/3\tan^{-1}(\sinh x)]$ with $c\not=0$, $\lambda=0.5$ and the parabola $V_0=-(2W_0^2/9+1)$.
Of course, the parabola is a subset of the condition (\ref{ptc}) (see Fig.~\ref{fig-spectra1}).

In the presence of momentum term $\Gamma$, we find that if $\lambda_0$ and $\Phi_0(x)$ satisfy $L_0\Phi_0(x)=\lambda_0\Phi_0(x)$, then the function $\Phi(x)$ via the invertible transformation $\Phi(x)=\Phi_0(x)e^{i\Gamma x}$
satisfies Eq.~(\ref{ls}) with  $\lambda=\lambda_0+\Gamma^2/2$.
Conversely, the results also hold, that is, for  the non-zero momentum coefficient $\Gamma$, the operator  $L_s$ admits entirely real (discrete) spectra provided that the parameters $V_0<0$ and $W_0$ satisfy the same condition (\ref{ptc}).

Moreover, we numerically study the domains of unbroken and broken $\PT$-symmetric phases of the operator $L_s$ with $\PT$-symmetric potential (\ref{ps}) for different frequencies on the $(V_0, W_0)$-space, which are the same as ones given by Eq.~(\ref{ptc}), that is,  the momentum coefficient $\Gamma$ can not change the domains of  unbroken/broken $\PT$-symmetric phase, but it has the effect on the eigenvalues for the given $\PT$-symmetric potential (see Fig.~\ref{fig-spectra1}).

\begin{figure}[!t]
	\begin{center}
		\hspace{-0.05in}{\scalebox{0.35}[0.35]{\includegraphics{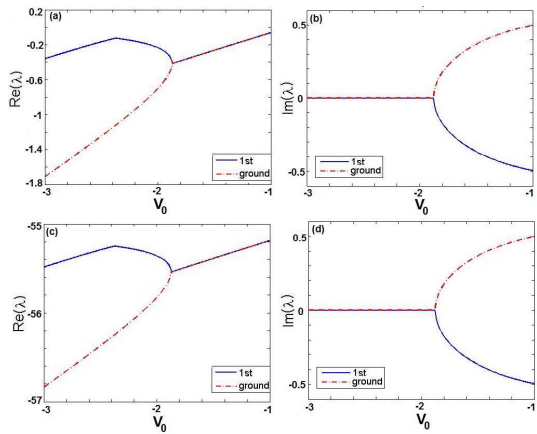}}}
				\end{center}
	\vspace{-0.15in}
	\caption{ (color online).  (a, c) Real and (b, d) imaginary parts of the eigenvalues $¦Ë$ [see Eq.~(\ref{ls})]
as functions of $V_0<0$ for the $\PT$-symmetric potential (\ref{ps}) at (a, b) $\Gamma=0.0375$ and (c, d) $\Gamma=10.5$. Other parameter is $W_0=2$.}		 \label{fig1i}
\end{figure}

 We choose the different momentum coefficients $\Gamma=0.0375$~\cite{rbecb} and $\Gamma=10.5$ as well as fix $W_0=2$ to numerically illustrate two lowest states corresponding to discrete spectra such that the spontaneous symmetry breaking occurs due to collision of the two lowest states as the amplitude $|V_0|$ of the potential decreases. Moreover, the absolute value of real part of the eigenvalue becomes large as the momentum coefficient increases (see Fig.~\ref{fig1i}). The non-smooth points are due to the order of energy levels (i.e, the real parts of eigenvalues) (see Figs.~\ref{fig1i}(a,c)).

\subsubsection{Nonlinear localized modes and stability}

We now turn to study nonlinear localized modes of Eq.~(\ref{ode}) in $\PT$-symmetric Scarff-II potential (\ref{ps}). We can find that Eq.~(\ref{ode}) possesses the unified exact bright solitons
\bee\label{solu}
 \phi(x)=\sqrt{\frac{1}{g}\left(\frac{2W_0^2}{9}+V_0+1\right)}\,{\rm sech}x \,e^{i\varphi(x)},
\ene
for both attractive ($g=1)$ and repulsive $(g=-1)$ nonlinearities, where the chemical potential $\mu$ is modulated by the frequency as
$\mu=(1+\Gamma^2)/2$, 
and the non-trivial phase is related to the amplitude $W_0$ of gain-and-loss distribution and momentum coefficient $\Gamma$, that is
$\varphi(x)=\Gamma x-\frac{2W_0}{3}\tan^{-1}(\sinh x)$.
The existence conditions of the bright solitons (\ref{solu}) are $V_0>-(2W_0^2/9+1)$ for $g=1$ and $V_0<-(2W_0^2/9+1)$ for $g=-1$ [see the regions inside ($g=-1)$ and outside $(g=1)$ the parabola $V_0=-(2W_0^2/9+1)$ in Fig.~\ref{fig-spectra1}]. Notice that for $\Gamma=0$, the chemical potential is a constant, $\mu=0.5$. When $\Gamma$ increases, $\mu$ also becomes large. Moreover, the momentum coefficient $\Gamma$ does not control the amplitude of nonlinear modes (\ref{solu}), but changes the chemical potential $\mu$ and phase $\varphi(x)$. It is easy to see that nonlinear modes (\ref{solu}) are also $\PT$-symmetric.

We can find that the parabola $V_0=-(2W_0^2/9+1)$ is tangent to the $\PT$-symmetric threshold lines $W_0=\pm(0.125-V_0)$ with two tangent points being $(V_0, W_0)=(-2.125, \pm 2.25)$, that is, the existence domain of bright solitons (\ref{solu}) for the repulsive case $g=-1$ is {\it completely} located in the region of unbroken $\PT$-symmetric phase, but the existence domain of bright solitons (\ref{solu}) for the attractive case $g=1$ contains both {\it partial domain} of unbroken $\PT$-symmetric phase and {\it entire domain} of broken $\PT$-symmetric phase (see Fig.~\ref{fig-spectra1}). Particularly, we find that the momentum coefficient $\Gamma$ can modulate both the phase $\varphi(x)$ and  chemical potential $\mu$, but is independent on the amplitudes of bright solitons (\ref{solu}). The amplitudes $V_0$ and $W_0$ of $V(x)$ and $W(x)$ can control the amplitudes of nonlinear modes (\ref{solu}) and  power since $P=\frac{2}{g}\left(\frac{2W_0^2}{9}+V_0+1\right)$ for nonlinear modes (\ref{solu}).

 \begin{figure}[!t]
 	\begin{center}
 	\vspace{0.05in}
 	\hspace{-0.05in}{\scalebox{0.55}[0.5]{\includegraphics{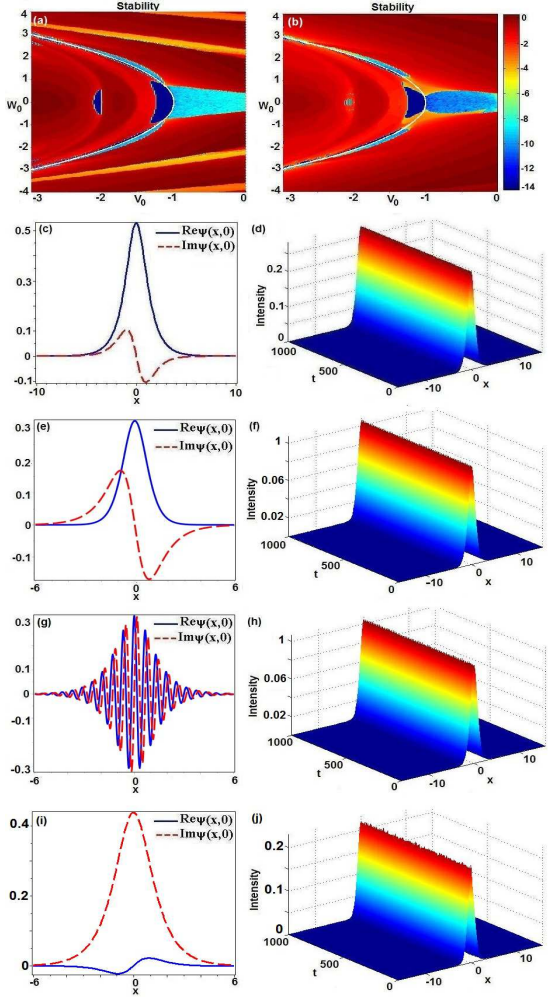}}}
 	\end{center}
 	\vspace{-0.15in} \caption{\small (color online)
  Linear stability [cf. Eq.~(\ref{st})] of nonlinear modes (\ref{solu}) for (a) $\Gamma=0.0375$ and (b) $\Gamma=10.5$ [the maximal absolute value of imaginary parts of the linearized eigenvalue $\delta$ in $(V_0, W_0)$-space (common logarithmic scale), similarly hereinafter], where the parabola is $V_0=-(2W_0^2/9+1)$. One-hump [(c) $\Gamma=0.0375,\, V_0=-0.8,\, W_0=0.6$ (unbroken linear $\PT$-symmetry), (e) $\Gamma=0.0375,\, V_0=-1.5,\,  W_0=1.65$ (broken linear $\PT$-symmetry), [(g) $\Gamma=10.5,\, V_0=-1.5,\, W_0=1.65$ (broken linear $\PT$-symmetry)], and (i) $\Gamma=0.0375,\, V_0=-1.2,\,  W_0=0.2$ (unbroken linear $\PT$-symmetry)] nonlinear modes (\ref{solu}). (d, f, h) Stable and (j) periodically varying propagation of the nonlinear modes (\ref{solu})  corresponding to the weakly perturbed initial conditions shown in (c, e, g), and (i) respectively. Other parameter is $g=1$.} \label{stability-g1}
 \end{figure}

 \begin{figure}[!t]
 	\begin{center}
 	\vspace{0.05in}
 	\hspace{-0.05in}{\scalebox{0.55}[0.5]{\includegraphics{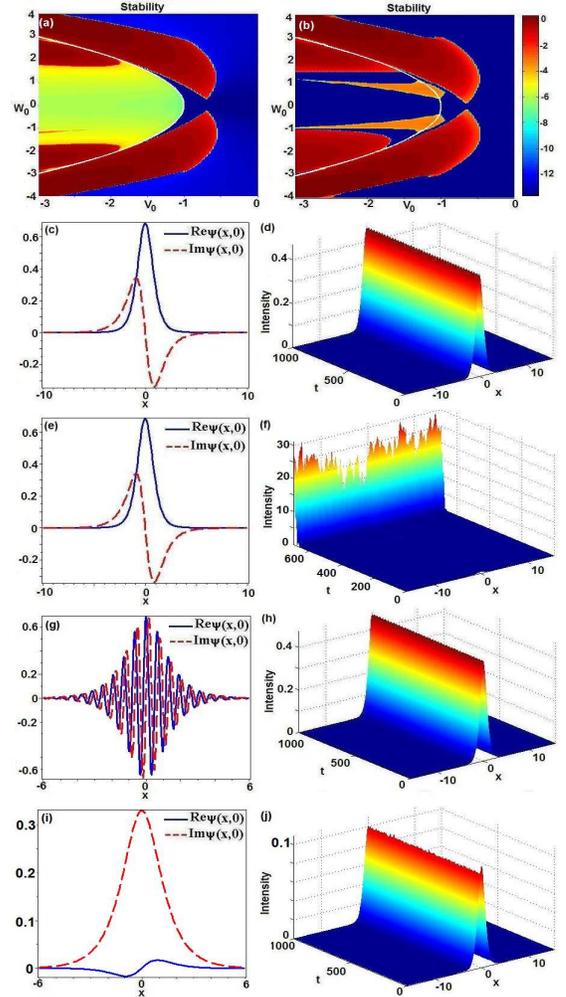}}}
 	\end{center}
 	\vspace{-0.15in} \caption{\small (color online).
   Linear stability [cf. Eq.~(\ref{st})] of nonlinear modes (\ref{solu}) for (a) $\Gamma=0.0375$ and (b) $\Gamma=10.5$, where the parabola is $V_0=-(2W_0^2/9+1)$.
 One-hump [(c) $\Gamma=0.0375,\, V_0=-2,\, W_0=1.5495$ (unbroken linear $\PT$-symmetry), (e) $\Gamma=0.0375,\, V_0=-2,\, W_0=1.5498$ (unbroken linear $\PT$-symmetry), (g) $\Gamma=10.5,\, V_0=-2,\, W_0=1.5498$ (unbroken linear $\PT$-symmetry), and  (i) $\Gamma=0.0375,\, V_0=-0.9,\,  W_0=0.2$ (unbroken linear $\PT$-symmetry)] nonlinear modes (\ref{solu}). (d, h) Stable, (f) unstable, and (j) periodically varying propagation of the nonlinear modes described by Eq. (\ref{solu}) subject to the weakly perturbed initial conditions shown in (c, g, e), and (i), respectively. Other parameter is $g=-1$.}
 	\label{stability-g-1}
 \end{figure}

 \begin{figure}[!t]
 	\begin{center}
 	\vspace{0.05in}
 	\hspace{-0.05in}{\scalebox{0.4}[0.4]{\includegraphics{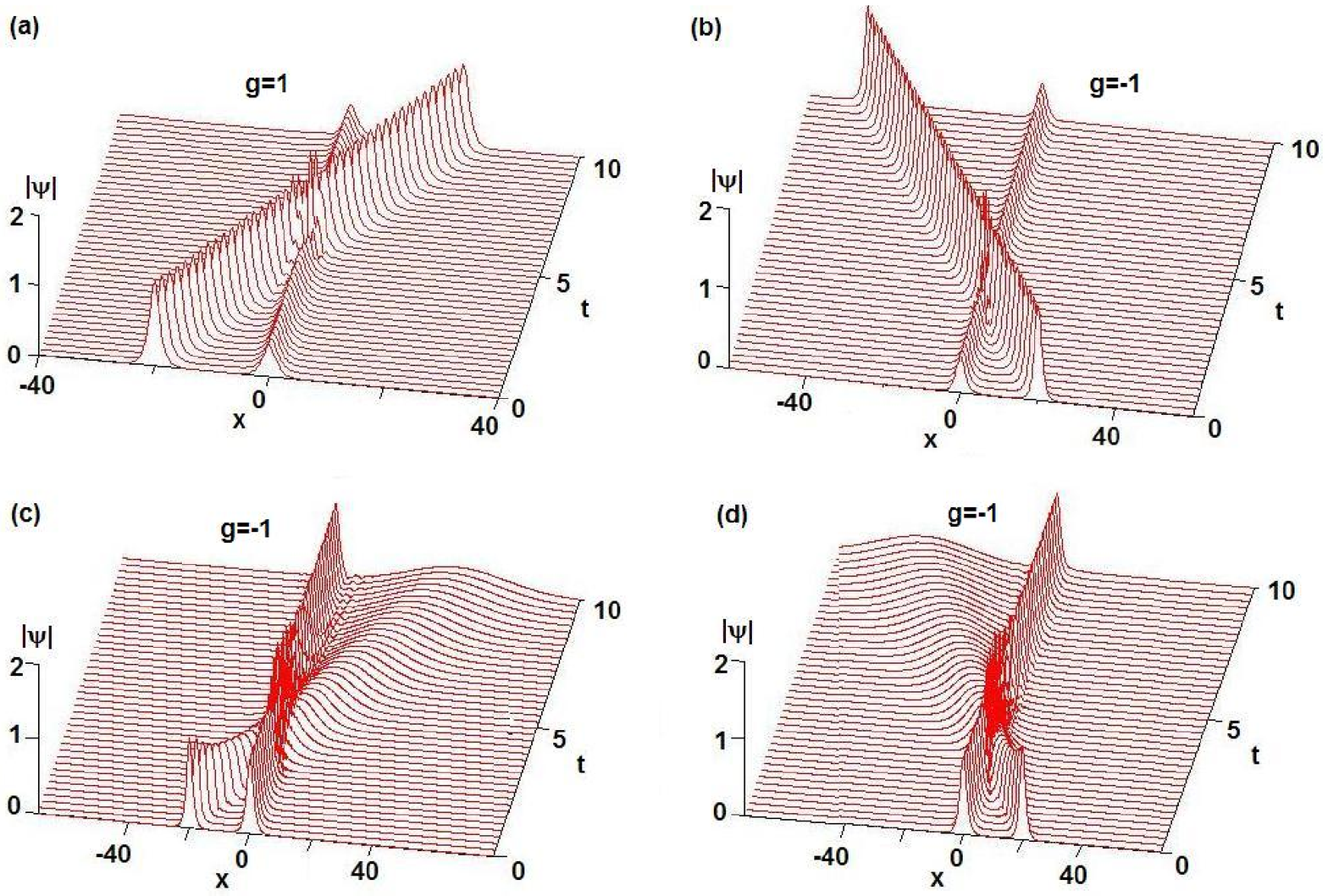}}}
 	\end{center}
 	\vspace{-0.15in} \caption{\small (color online). The interactions of two solitons in Eq.~(\ref{nls}). (a) the solution (\ref{solu}) with the wave $1.2\,{\rm sech}[1.2(x+20)]e^{4ix}$ with $\Gamma=0.0375,\, V_0=-0.8, W_0=0.1, g=1$, (b) the solution (\ref{solu}) with the wave $1.2\,{\rm sech}[1.2(x-20)]e^{4ix}$ with $\Gamma=10.5,\, V_0=-0.8, W_0=0.1, g=1$, (c)  the solution (\ref{solu}) with the wave $1.2\,{\rm sech}[1.2(x+20)]e^{6ix}$ with $\Gamma=0.0375,\, V_0=-2, W_0=0.1, g=-1$, (d) The solution (\ref{solu}) with the wave $1.2\,{\rm sech}[1.2(x-20)]e^{4ix}$ with
 $\Gamma=10.5,\, V_0=-2, W_0=0.1, g=-1$. }
 	\label{osolu}
 \end{figure}

For the chosen momentum coefficients $\Gamma=0.0375,\, 10.5$, Figs.~\ref{stability-g1}(a, b) for $g=1$ and Figs.~\ref{stability-g-1}(a, b) for $g=-1$ display  maximal absolute values of imaginary parts of linearized eigenvalues $\delta$ related to solutions (\ref{solu}) as functions of $V_0$ and $W_0$ [cf. Eq.~\ref{st})], which illustrate the linear stable (the dark blue) and  unstable (other) regions.
In the following, we numerically check the robustness of nonlinear modes (\ref{solu}) for both attractive and repulsive cases via the direct propagation of initially stationary state (\ref{solu}) with a noise $1\%$.

 For the attractive case $g=1$, Fig.~\ref{stability-g1} illustrates the stable and unstable situations for different parameters $V_0,\, W_0$, and $\Gamma$. For $\Gamma=0.0375$ and $V_0=-0.8,\, W_0=0.6$ belonging to the region of unbroken linear $\PT$-symmetric phase of the operator $L_s$ [cf. Eq.~(\ref{ls})], the nonlinear mode is stable (see Figs.~\ref{stability-g1}(c, d)), which can be observed in the relevant experiments. If we fix $\Gamma=0.0375,\, V_0=-0.8$ and change $W_0=0.8$, which even if corresponds to the domain of unbroken linear $\PT$-symmetric phase, but the nonlinear mode becomes unstable. Similarly, if we fix $V_0=-0.8,\, W_0=0.6$ and change $\Gamma=10.5$, which also corresponds to the domain of unbroken linear $\PT$-symmetric phase, then we find that the nonlinear mode also is unstable. If we choose $V_0=-1.5,\, W_0=1.65$ belonging to the domain of broken linear $\PT$-symmetric phase [cf. Eq.~(\ref{ls})], then we find the stable nonlinear modes for $\Gamma=0.0375$ and $\Gamma=10.5$ [see Figs.~\ref{stability-g1}(f, h)].

Notice that for the attractive case $g=1$, exact nonlinear modes (\ref{solu}) of Eq.~(\ref{ode}) exist outside the parabola $V_0=-(2W_0^2/9+1)$ [see the white parabola in Figs.~\ref{stability-g1}(a, b)]. If we choose $V_0=-1.5,\, W_0=0.2$ (located inside the parabola), in which nonlinear modes (\ref{solu}) become $\phi_+(x)=i\sqrt{41}/15\,{\rm sech}x \exp[i(\Gamma x-2/15\tan^{-1}(\sinh x))]$ (see Fig.~\ref{stability-g1}(i), where the real part is a odd function and imaginary part is an even function, which differs from other cases [cf. Figs.~\ref{stability-g1}(c, e, g)]), which does not satisfies Eq.~(\ref{ode}) with $\PT$-symmetric Scarff-II potential (\ref{ps}). But we still use it as an initial solution with a noise perturbation of order $1\%$ to make numerical simulations such that we surprisedly find the initial mode $\phi_+(x)$ can be excited to a stable nonlinear mode, which exhibits the weak oscillations (breather-like behavior) (see Fig.~\ref{stability-g1}j). This may be due to the strong gain-and-loss distribution and non-exact initial condition and enlarges the modulation scope of parameters in the experiments.

For the repulsive case $g=-1$, Fig.~\ref{stability-g-1} illustrates the stable and unstable situations for different parameters $V_0,\, W_0$ and $\Gamma$. For $\Gamma=0.0375$ and $V_0=-2,\, W_0=1.5495$ related to the unbroken linear $\PT$-symmetric phase of $L_s$, the nonlinear mode is stable (see Figs.~\ref{stability-g-1}(c, d)). If we fix $\Gamma=0.0375,\, V_0=-2$ and change $W_0$ a little into $W_0=1.5498$ related to unbroken linear $\PT$-symmetric phase of $L_s$, then the nonlinear mode becomes unstable (see Figs.~\ref{stability-g-1}(e, f)), whereas we fix $W_0=1.5498,\, V_0=-2$ and change $\Gamma=10.5$ related to the unbroken linear $\PT$-symmetric phase of $L_s$ such that the nonlinear mode becomes stable again (see Figs.~\ref{stability-g-1}(g, h)).

Notice that for the repulsive case $g=-1$, nonlinear mode (\ref{solu}) exists inside the parabola $V_0=-(2W_0^2/9+1)$ (see the parabola (white parabola) in Figs.~\ref{stability-g-1}(a, b)). If we choose $V_0=-0.9,\, W_0=0.2$ (located outside the parabola), in which nonlinear mode (\ref{solu}) becomes $\phi_{-}(x)=i\sqrt{53/450}\,{\rm sech}x \exp[i(\Gamma x-2/15\tan^{-1}(\sinh x))]$ (see Fig.~\ref{stability-g-1}(i)), which does not satisfy Eq.~(\ref{ode}) with $\PT$-symmetric Scarff-II potential (\ref{ps}). But we still use it as an initial solution with a noise $1\%$ to make numerical simulations such that we surprisedly find the initial mode $\phi_{-}(x)$ can be excited to a stable nonlinear mode and we observe the phenomenon that there is a small  bulge  at the beginning, and then it decreases  such that the weak oscillatory (breather-like behavior) situation is generated (see Fig.~\ref{stability-g-1}(j)).

Moreover, we also study the interactions of two bright solitons in the $\PT$-symmetric potential. For the attractive case  $g=1$ and $V_0=-0.8,\, W_0=0.1$, we consider the initial condition $\psi(x,0)=\phi(x)+1.2\,{\rm sech}[1.2(x+20)]e^{4ix}$ with $\phi(x)$ given by Eq.~(\ref{solu}) and $\Gamma=0.0375$ such that the elastic interaction is generated (see Fig.~\ref{osolu}(a)). When $\Gamma$ becomes large, e.g., $\Gamma=10.5$, we consider the initial condition $\psi(x,0)=\phi(x)+1.2\,{\rm sech}[1.2(x-20)]e^{4ix}$ with $\phi(x)$ given by Eq.~(\ref{solu}) (see Fig.~\ref{osolu}(b)). Similarly, for the repulsive case  $g=-1$ and $V_0=-2,\, W_0=0.1$, we consider the initial condition $\psi(x,0)=\phi(x)+1.2\,{\rm sech}[1.2(x+20)]e^{6ix}$ with $\phi(x)$ given by Eq.~(\ref{solu}) and $\Gamma=0.0375$ such that the semi-elastic interaction is generated in which the shapes of exact nonlinear mode are almost same before and after interaction, but the shapes of the other mode are changed before and after interaction (see Fig.~\ref{osolu}(c)). When $\Gamma$ becomes big, e.g., $\Gamma=10.5$, we consider the initial condition $\psi(x,0)=\phi(x)+1.2\,{\rm sech}[1.2(x-20)]e^{4ix}$ with $\phi(x)$ given by Eq.~(\ref{solu}) such that we find
 the similar semi-elastic interaction (see Fig.~\ref{osolu}(d)).

\begin{figure}[!t]
	\begin{center}
		\hspace{-0.05in}{\scalebox{0.2}[0.15]{\includegraphics{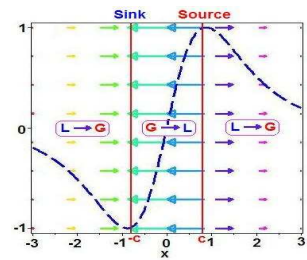}}}
				\end{center}
	\vspace{-0.2in}
	\caption{ (color online). The transverse power-flow (\ref{svector1}) related to nonlinear modes (\ref{solu}) and gain-and-loss curves $W(x)$ given by Eq.~(\ref{ps}) (dashed lines) with $W_0=2,\, \Gamma=1,\, g=1, V_0=-1$. $c={\rm sech}^{-1}(3/4)$, `L'=`Loss', `G'=`Gain'.}	 \label{fig-vector}
\end{figure}

\subsubsection{The varying transverse power-flow of nonlinear modes}

To more understand the properties of stationary nonlinear mode (\ref{solu}), we check its transverse power-flow or `Poynting vector',
which arises from the nontrivial phase structure of the nonlinear mode and is
\bee \label{svector1} \begin{array}{rl}
 S(x)&=\frac{i}{2}(\phi\phi_x^{*}-\phi_x\phi^{*}) \vspace{0.05in} \cr
  & =\left[\dfrac{1}{g}\!\left(\dfrac{2W_0^2}{9}\!+\!V_0\!+\!1\right){\rm sech}^2x\right]\!\!\!\left(\Gamma\!-\!\dfrac{2W_0}{3}{\rm sech}x\right),
\end{array} \,\,\,\quad
\ene
For $\Gamma=0$, the ${\rm sgn}(S)=-{\rm sgn}(W_0)$, that is, the power always flows from the gain toward loss (see Refs.~\cite{ptsf, yanpre15}), but in the presence of momentum term $\Gamma\not=0$, the sign of $S$ is not always positive or negative definite and is dependent on both parameters $W_0$ and $\Gamma$, and even space position $x$. As a consequence, these four results are generated in Table~\ref{table}. We discuss them as follows:

  i) For the Case 1, the power always flows from the loss toward the gain.

ii)  For the Case 2, the the power flowing directions  are complicated in different positions. For any $0<\Gamma<2W_0/3$ with $W_0>0$,
 the power always flows from the loss toward the gain for $|x|\geq {\rm sech}^{-1}(3\Gamma/(2W_0))$ and from the gain toward the loss for $|x|< {\rm sech}^{-1}(3\Gamma/(2W_0))$ (see Fig.~\ref{fig-vector}).

  iii)  For the Case 3, we find that the power always flows from the gain toward the loss, which displays that the momentum operator $i\Gamma\partial_x$ is the same (or weak) action as the gain-and-loss term $W(x)$ ($\Gamma\not=0$) or has no effect on gain-and-loss term (i.e., $\Gamma=0$).

  iv) For the Case 4, i.e., without the gain-and-loss term $W(x)\equiv 0$, the  momentum  operator $i\Gamma\partial_x$ has the similar action on power flowing direction such that the power always flows in one direction.


\begin{table}[!t]
\vspace{-0.1in}
\caption{\small The signs of $S(x)$ (the power flowing direction) is related to $W_0$, $\Gamma$, and even space (cf. Eq.~(\ref{svector1})). \vspace{0.05in}}
\begin{tabular}{ccll} \hline\hline \\ [-2.0ex]
 Case &    $W_0$   & \qquad $\Gamma$   &  \quad $S(x)$ \\ [1.0ex] \hline \\ [-2.0ex]
1 &  $W_0>0$  \,\,  & $\Gamma\geq \frac{2W_0}{3}$  & $S\geq 0$ \,\, {\rm for~all~} $x$ \\ [1.0ex] \hline \\ [-2.0ex]
2&  $W_0>0$ \,\, & $0\!<\!\Gamma\!<\!\frac{2W_0}{3} $ \qquad &
               $\begin{cases}S\geq 0, \,\,  \,\, |x|\!\geq\! {\rm sech}^{-1}\!\left(\frac{3\Gamma}{2W_0}\right) \vspace{0.02in}\\
                               S< 0, \,\, \,\ otherwise
                                                               \end{cases}$     \\ [1.0ex] \hline \\ [-2.0ex]
3&  $W_0<0$ \,\,  & $\Gamma >0$  & $S> 0$ \,\, {\rm for~all~} $x$ \\ [1.0ex] \hline \\ [-2.0ex]
4& $W_0=0$ \,\, & $\Gamma > 0$  & $S >0$ \,\, {\rm for~all~} $x$     \\ [1.0ex] \hline\hline
\end{tabular}
\label{table}
\end{table}

\subsection{$\PT$-symmetric $\alpha$-power-law Scarff-II potential}

We here consider the $\PT$-symmetric $\alpha$-power-law Scarff-II potential
\bee  \nonumber
 V_{\alpha}(x)\!=\!v_1{\rm sech}^2x\!+\!v_2{\rm sech}^{2\alpha}x,
 W_{\alpha}(x)\!=\!W_0{\rm sech}^{\alpha}x\tanh x, \\
\label{gps}
\ene
where $\alpha>0$, $v_1=-\alpha(\alpha+1)/2$, $v_2$ and $W_0$ are both real constants. For the case $\alpha=1$, the potential $V_{\alpha}(x)$ and $W_{\alpha}(x)$ become the well-known Sacrff-II potential $V_1(x)=(v_1+v_2){\rm sech}^2x$ and $W_1(x)=W_0{\rm sech}x\tanh x$~\cite{scarff}.

For the different parameters $\alpha>0$ and $v_2$, the potential $ V_{\alpha}(x)$ displays abundant well structures (see Figs.~\ref{stability05} and \ref{stability2}).  For the potential
 (\ref{gps}), we have the bright solitons of Eq.~(\ref{ode})
\bee\label{solu3}
\phi_{\alpha}(x)=\sqrt{\frac{1}{g}\left(\frac{2W_0^2}{9\alpha^2}+v_2\right)}\,\,{\rm sech}^{\alpha}x\,e^{i\varphi_{\alpha}(x)}
\ene
for $g=\pm 1$, where $g(v_2+2W_0^2/(9\alpha^2))>0$,  the chemical potential  is $\mu=(\alpha^2+\Gamma^2)/2$ and the phase is
\bee
 \varphi_{\alpha}(x)=\Gamma x-\frac{2W_0}{3\alpha}\int^x_0{\rm sech}^{\alpha}s ds,
\ene

Notice that when $v_2=0$, we know that the solution (\ref{solu3}) is only for the case $g=1$. For $0<\alpha<1$ the wave widths of nonlinear modes (\ref{solu3}) become larger than one with $\alpha=1$, whereas for $\alpha>1$, they are smaller than ones with  with $\alpha=1$.
  In the following we numerically study the linear stability of nonlinear modes (\ref{solu3}) for two cases $0<\alpha<1$ (e.g., $\alpha=0.5$) and
  $\alpha>1$ (e.g., $\alpha=2$).

 \subsubsection{Stability of nonlinear modes}

In the following, we check the robustness of nonlinear modes (\ref{solu3}) for both attractive and repulsive cases by numerical simulations via the direct propagation of the initially stationary state in
Eq.~(\ref{solu3}) with a noise perturbation of order $1\%$.

\begin{figure}[!t]
 	\begin{center}
 	\vspace{0.05in}
 	\hspace{-0.05in}{\scalebox{0.48}[0.4]{\includegraphics{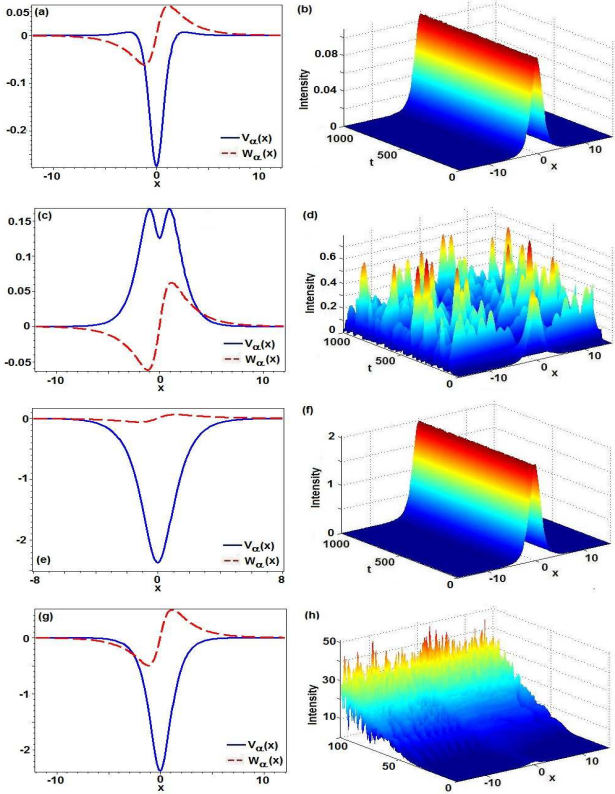}}}
 	\end{center}
 	\vspace{-0.15in} \caption{\small (color online). $\PT$-symmetric potential given by Eq.~(\ref{gps}) with (a) $v_2=W_0=0.1,\, g=1$, (c) $v_2=0.5, W_0=0.1,\, g=1$, (e)
  $v_2=-2,\, W_0=0.1, g=-1$, and (g) $v_2=-2, W_0=0.8, g=-1$. (b, f) Stable and (d, h) unstable propagations of nonlinear modes described by Eq.~(\ref{solu3}) subject to the $\PT$-symmetric potentials (a, e) and (c, g), respectively. Other parameters are $\Gamma=1$ and $\alpha=0.5$.}
 	\label{stability05}
 \end{figure}

We fix $\alpha=0.5$ and $\Gamma=1$. For the attractive case $g=1$, we have the stable modes (Fig.~\ref{stability05}(b)) for $v_2=W_0=0.1$, in which the potential is single-well-like (see Fig.~\ref{stability05}(a)). If we increase $v_2$, e.g., $v_2=0.5$, in which the potential becomes the single-well with double-hump (see Fig.~\ref{stability05}(c)) such that we have the unstable modes (Fig.~\ref{stability05}(d)). For the repulsive case $g=-1$, we fix $v_2=-2$ and change $W_0$ such that we have the stable (Fig.~\ref{stability05}(f)) and unstable (Fig.~\ref{stability05}(h)) for $W_0=0.1$ and $W_0=0.8$, respectively. The two cases only lead to the single-well potential (Figs.~\ref{stability05}(e, g)).

\begin{figure}[!t]
 	\begin{center}
 	\vspace{0.05in}
 	\hspace{-0.05in}{\scalebox{0.48}[0.4]{\includegraphics{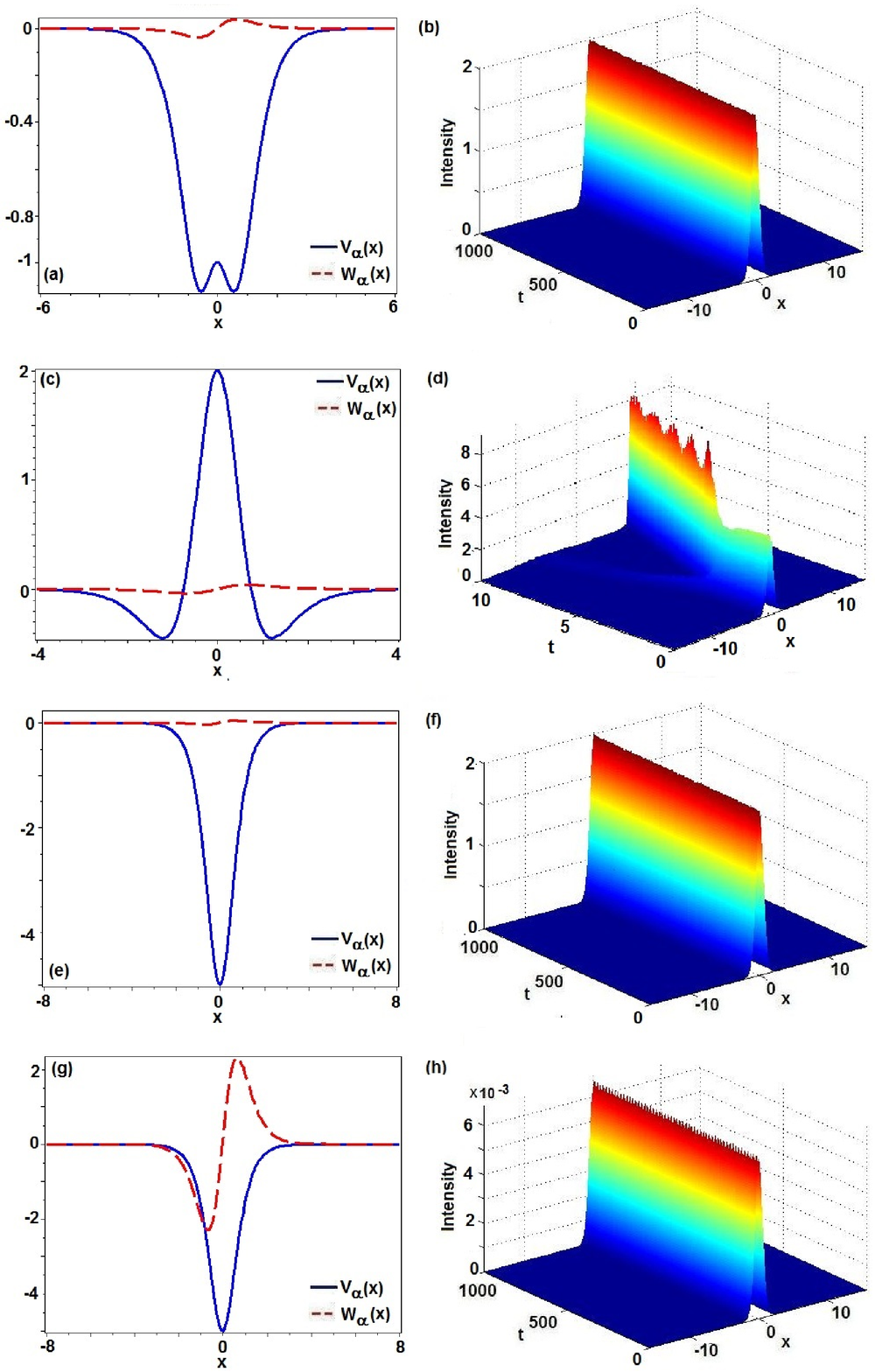}}}
 	\end{center}
 	\vspace{-0.15in} \caption{\small (color online). $\PT$-symmetric potential given by Eq.~(\ref{gps}) with (a) $v_2=2, W_0=0.1,\, g=1$, (c)
 $v_2=5, W_0=0.1,\, g=1$, (e) $v_2=-2,\, W_0=0.1, g=-1$, and (g) $v_2=-2, W_0=5.99, g=-1$. (b, f, h) Stable and (d) unstable propagations of nonlinear modes described by Eq.~(\ref{solu3}) subject to the $\PT$-symmetric potentials (a, e, g), and (c), respectively.
  Other parameters are  $\Gamma=1$ and $\alpha=2$.}
 	\label{stability2}
 \end{figure}

We fix $\alpha=2,\, \Gamma=1$. For the attractive case $g=1$, we have the stable modes (Fig.~\ref{stability2}(b)) for $v_2=2,\, W_0=0.1$, in which the potential is double-well (Fig.~\ref{stability2}(a)). If we increase $v_2$ as $v_2=5$ in which the potential becomes the double-well-like with higher center hump (Fig.~\ref{stability2}(c)) such that we have the unstable modes (Fig.~\ref{stability2}(d)). For the repulsive case $g=-1$, we fix $v_2=-2$ and change $W_0$ such that we have the stable (Fig.~\ref{stability2}(f, h)) for both $W_0=0.1$ and $W_0=5.99$, respectively. In fact, if we fix $v_2=-2$, then the nonlinear mode (\ref{solu3}) is still stable for $|W_0|<6$ and $g=-1$.  The two cases only lead to the single-well potential (Figs.~\ref{stability2}(e, g)).

\subsubsection{The  varying transverse power-flow}

To more understand the properties of stationary nonlinear mode (\ref{solu3}), we check its transverse power-flow or `Poynting vector' in the form
\bee
 S(x)\!=\!\left[\frac{1}{g}\left(v_2\!+\!\frac{2W_0^2}{9\alpha^2}\right)\!{\rm sech}^{2\alpha}x\right]\!\left(\Gamma\!-\!\dfrac{2W_0}{3\alpha}{\rm sech}^{\alpha}x\right),\,\,\,
\ene
which denotes that the sign of $S$ is not positive or negative definite and is dependent on both parameters $W_0,\, \alpha$ and $\Gamma$, and even space position $x$. These four results are listed in Table~\ref{table-p}. We discuss them as follows:

  i) For the Case 1, the power always flows from the loss toward the gain.

ii)  For the Case 2, the the power flowing directions  are complicated. For any $0<\Gamma<2W_0/(3\alpha)$ with $W_0>0$,
 the power always flows from the loss toward the gain for $|x|\geq {\rm sech}^{-1}(\sqrt[\alpha]{3\alpha\Gamma/(2W_0)})$ and from the gain toward the loss for $|x|< {\rm sech}^{-1}(\sqrt[\alpha]{3\alpha\Gamma/(2W_0)})$.

  iii)  For the Case 3, we find that the power always flows from the gain toward the loss, which displays that the  momentum  operator $i\Gamma\partial_x$ is the same (or weak) action as the gain-and-loss term $W(x)$ ($\Gamma\not=0$) or has no effect on gain-and-loss term (i.e., $\Gamma=0$).

iv) For the Case 4, i.e., without the gain-and-loss term $W(x)\equiv 0$, the  momentum  operator $i\Gamma\partial_x$ has the similar action on power flowing direction such that the power always flows in one direction.


\begin{table}[!t]
\vspace{-0.1in}
\caption{\small The sign of $S(x)$ (the power flowing direction) is related to $W_0$, $\Gamma$, and even space. \vspace{0.05in}}
\begin{tabular}{ccll} \hline\hline \\ [-2.0ex]
 Case &    $W_0$   & \qquad $\Gamma$   &  \quad $S(x)$ \\ [1.0ex] \hline \\ [-2.0ex]
1 &  $W_0>0$  \,\,  & $\Gamma\geq \frac{2W_0}{3\alpha}$  & $S\geq 0$ \,\, {\rm for~all~} $x$ \\ [1.0ex] \hline \\ [-2.0ex]
2&  $W_0>0$ \,\, & $0\!<\!\Gamma\!<\!\frac{2W_0}{3\alpha} $ \qquad &
               $\begin{cases} S\geq 0, \,\, \,\, |x|\!\geq\! {\rm sech}^{-1}\!\left(\!\!\!\sqrt[\alpha]{\frac{3\alpha\Gamma}{2W_0}}\right) \vspace{0.05in}\\
                               S< 0, \,\, \,\ otherwise     \end{cases}$     \\ [1.0ex] \hline \\ [-2.0ex]
3&  $W_0<0$ \,\,  & $\Gamma >0$  & $S> 0$ \,\, {\rm for~all~} $x$ \\ [1.0ex] \hline \\ [-2.0ex]
4& $W_0=0$ \,\, & $\Gamma > 0$  & $S >0$ \,\, {\rm for~all~} $x$     \\ [1.0ex] \hline\hline
\end{tabular}
\label{table-p}
\end{table}

\subsection{$\PT$-symmetric harmonic-Gaussian potential}

We here consider another physically relevant case of the
parabolic (harmonic) potential
\bee\label{ps2a}
V(x)=\frac{1}{2}\omega^2x^2,
\ene
and the family of Hermite-Gaussian type of gain-and-loss distributions~\cite{harm1, yanpra15}
\bee\label{ps2b}
W_n(x)\!=\!\sigma[\omega xH_{n}(\sqrt{\omega}x)\!-\!2n\sqrt{\omega}H_{n-1}(\sqrt{\omega}x)]e^{-\omega x^2/2},\,\,\,\,\,\,\,
\ene
where the frequency $\omega>0$, $\sigma$ is a real amplitude, and $H_n(x)=(-1)^ne^{x^2}(d^ne^{-x^2})/(dx^n)$ denotes the Hermite polynomial with $n$ is a non-negative integer and $H_n(x)\equiv 0$ with $n<0$. For the non-negative even numbers $n=0,2,4,...$, the complex potential
$V(x)+iW_n(x)$ are all just $\PT$-symmetric.  For the non-negative odd numbers $n=1,3, 5,...$, the complex potential
$V(x)+iW_n(x)$ are all non-$\PT$-symmetric since $W_n(x)$ are all even functions.

In what follows, we mainly study both $\PT$-symmetric potentials for $n=0, 2$ and non-$\PT$-symmetric potential for $n=1$.

\subsubsection{Linear eigenvalue problem}

Here we consider the rotating operator $L_s$ with $\PT$-symmetric potentials with $V(x)$ (\ref{ps2a}) and $W_{0,2}(x)$ (\ref{ps2b}), which are given by
 \bee \label{psi0}
   W_0(x)=\sigma\omega xe^{-\omega x^2/2},
   \ene
\bee \label{psi2}
 W_2(x)=2\sigma\omega (2\omega x^3-5x)e^{-\omega x^2/2},
 \ene

\begin{figure}[!t]
 	\begin{center}
 	\vspace{0.05in}
 	\hspace{-0.05in}{\scalebox{0.4}[0.4]{\includegraphics{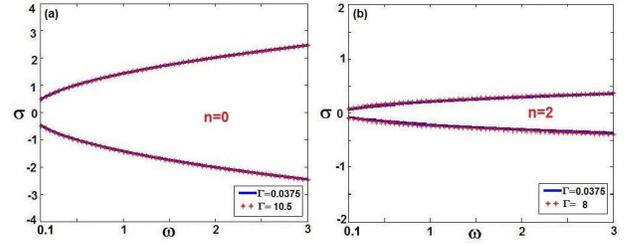}}}
 	\end{center}
 	\vspace{-0.15in} \caption{\small (color online). The profiles of phase transitions for the linear operator $L_s$ (\ref{ls}) with $\PT$-symmetric harmonic-Gaussian potentials with $V(x)$ (\ref{ps2a}) and $W_0$ (\ref{psi0}) and $W_2$ (\ref{psi2}). The unbroken (broken) $\PT$-symmetric phase is in the domain between (outside) two symmetric phase breaking curves for (a) $n=0$,\, $\Gamma=0.0375, 10.5$, and (b)  $n=2$,\, $\Gamma=0.0375, 8$. }
 	\label{fig-spectra1-2r}
 \end{figure}

Figs.~\ref{fig-spectra1-2r}(a) and (b) numerically exhibit the domains of the unbroken and broken $\PT$-symmetric phases of the rotating operator $L_s$ with $\PT$-symmetric potentials with $V(x)$ (\ref{ps2a}) and $W_0(x)$ (\ref{psi0}) or $W_2(x)$ (\ref{psi2}) for the different frequency $\Gamma$ on the $(\omega,\sigma)$-space, respectively. When $n$ increases, the domain of unbroken $\PT$-symmetric phase becomes small, whose main reason is due to the effect of the gain-and-loss distribution $W_n(x)$ as $n$ increases.

For the case $n=1$, we have the gain-and-loss distribution in an even function
\bee \label{psi1}
 W_1(x)=2\sigma\sqrt{\omega}(\omega x^2-1)e^{-\omega x^2/2},
 \ene
such that the complex potential $V(x)+iW_1(x)$ are non-$\PT$-symmetric. By using the numerical calculation, for the fixed $\Gamma=0.0375, 10.5$,  we can not almost find the domain of the unbroken phase of the rotating operator $L_s$ with non-$\PT$-symmetric potentials with $V(x)$ (\ref{ps2a}) and $W_1(x)$ (\ref{psi1}) in the same domain of $(\omega,\sigma)$-space $\{(\omega,\sigma)| 0<|\sigma|<4, 0<\omega<3\}$. This may imply that the $\PT$-symmetric potential indeed plays an important role
in the study of unbroken domains of the linear eigenvalue problems with complex potentials. But this result does not imply that non-$\PT$-symmetric potentials are not interesting.
We still find the stable nonlinear modes in the non-$\PT$-symmetric potential (see the following Sec.III-C-2.3).

\subsubsection{Nonlinear modes and stability}

For the above-mentioned $\PT$-symmetric potential (\ref{ps2a}) and (\ref{ps2b}), we have exact multi-hump bright solitons of  Eq.~(\ref{ode}) for the attractive case $g=1$
\bee \label{solu2}
\phi_n(x)=\frac{\sigma}{3}\sqrt{\frac{2}{g}}H_n(\sqrt{\omega}x)e^{-\omega x^2/2}e^{i\varphi_n(x)},
\ene
where the chemical potential $\mu=\frac12[\Gamma^2-(2n+1)\omega]$, and the phase is
\bee
 \varphi_n(x)=\Gamma x-\frac{2\sigma}{3}\int_0^x H_n(\sqrt{\omega}s)e^{-\omega s^2/2}ds
\ene

In the following we numerically study the linear stability of nonlinear modes (\ref{solu2}) for $n=0,1,2$.

\vspace{0.2in} \centerline{\it 2.1\, $\PT$-symmetric nonlinear modes $(n=0)$}
\vspace{0.1in}

 \begin{figure}[!t]
 	\begin{center}
 	\vspace{0.05in}
 	\hspace{-0.05in}{\scalebox{0.4}[0.35]{\includegraphics{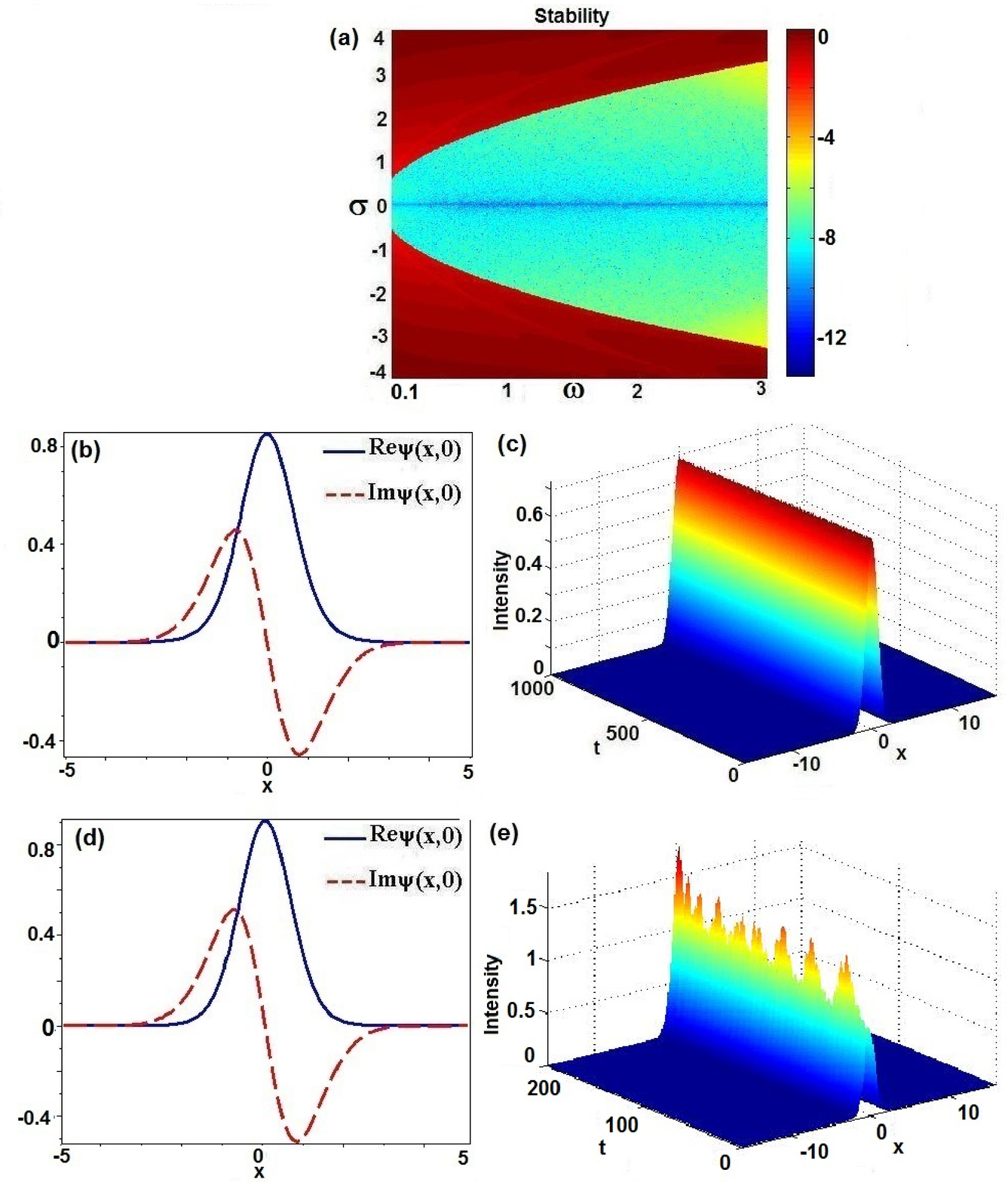}}}
 	\end{center}
 	\vspace{-0.15in} \caption{\small (color online).  (a) Linear stability [cf. Eq.~(\ref{st})] of nonlinear modes (\ref{solu2}).
One-hump nonlinear modes (\ref{solu2}) for (b) $\sigma=1.8$ (broken linear $\PT$-symmetry) and (d) $\sigma=1.92$ (broken linear $\PT$-symmetry). (c) Stable and (e) unstable propagation of the nonlinear modes (\ref{solu2}) corresponding to the weakly perturbed initial conditions shown in (b) and (d), respectively. Other parameters are $g= \omega=1,\, \Gamma=0.0375, \, n=0$. }
 	\label{stability-W0-0.0375}
 \end{figure}

 \begin{figure}[!t]
 	\begin{center}
 	\vspace{0.05in}
 	\hspace{-0.05in}{\scalebox{0.4}[0.35]{\includegraphics{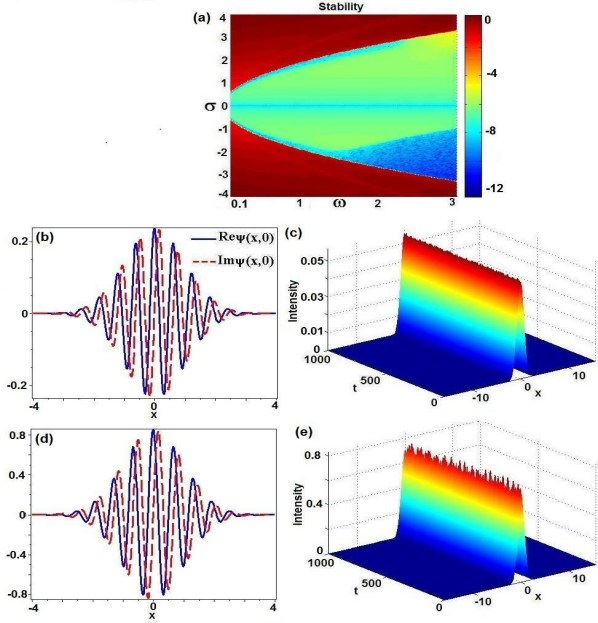}}}
 	\end{center}
 	\vspace{-0.15in} \caption{\small (color online).   (a)  Linear stability [cf. Eq.~(\ref{st})] of nonlinear modes (\ref{solu2}). One-hump nonlinear modes (\ref{solu2}) for (b) $\sigma=0.5$ (unbroken linear $\PT$-symmetry) and (d) $\sigma=1.8$ (broken linear $\PT$-symmetry). (c) Stable and (e) periodically varying propagation of the nonlinear modes (\ref{solu2}) corresponding to the weakly perturbed initial conditions shown in (b) and (d), respectively. Other parameters are $g= \omega=1,\, \Gamma=10.5,\, n=0$.}
 	\label{stability-W0-10.5}
 \end{figure}

 For the case  $n=0$, we firstly fix the  momentum coefficient $\Gamma=0.0375$ and give the domain of linear stability [cf. Eq.~\ref{st})] of nonlinear mode (\ref{solu2}) in the $(\omega, \sigma)$-space (see  Fig.~\ref{stability-W0-0.0375}(a)). Moreover, we present the wave propagations of nonlinear modes exhibited in Figs.~\ref{stability-W0-0.0375}(c, e) via the direct propagation of the initially stationary state in Eq.~(\ref{solu2}) with a noise perturbation of order about $1\%$ for special parameters $\omega$ and $\sigma$. In Fig.~\ref{stability-W0-0.0375}(c) we show that the nonlinear mode is stable even if the linear $\PT$-symmetric phase is broken. That is, the nonlinearity can excite the broken $\PT$-symmetric phase to the unbroken $\PT$-symmetric phase for some parameters.
If we increase the amplitude $\sigma$ of the gain-and-loss distribution a little bit such that we have the unstable nonlinear mode (see Fig.~\ref{stability-W0-0.0375}(d)).

We now change the momentum coefficient as $\Gamma=10.5$ and give the domain of linear stability [cf. Eq.~\ref{st})] of nonlinear mode (\ref{solu2}) in the $(\omega, \sigma)$-space (see  Fig.~\ref{stability-W0-10.5}(a)). Moreover, we present the wave propagations of nonlinear modes is exhibited in Fig.~\ref{stability-W0-10.5} via the direct propagation of the initially stationary state in Eq.~(\ref{solu2}) with a noise perturbation of order $1\%$. In Fig.~\ref{stability-W0-10.5}(c) we show that the nonlinear mode is stable where the linear $\PT$-symmetric phase is unbroken. If we increase the amplitude $\sigma$ of the gain-and-loss distribution in which the linear $\PT$-symmetric phase becomes broken such that we also have the stable nonlinear mode (see Fig.~\ref{stability-W0-10.5}(e)), but we observe the evident oscillations (breather-like behavior).

\vspace{0.2in} \centerline{\it 2.2 \, $\PT$-symmetric nonlinear modes $(n=2)$}
\vspace{0.1in}

 \begin{figure}[!t]
 	\begin{center}
 	\vspace{0.05in}
 	\hspace{-0.05in}{\scalebox{0.4}[0.35]{\includegraphics{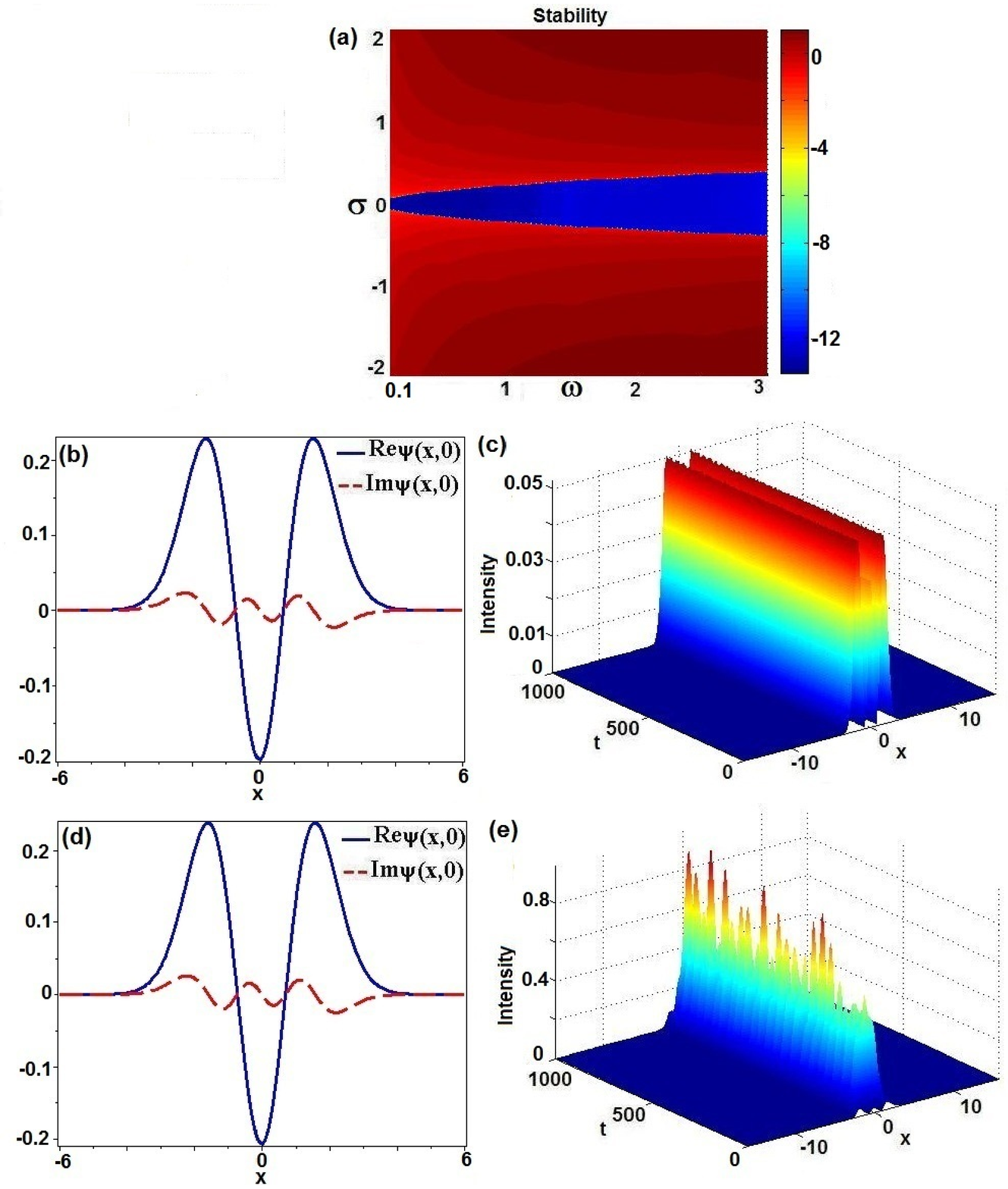}}}
 	\end{center}
 	\vspace{-0.15in} \caption{\small (color online). (a) Linear stability  [cf. Eq.~(\ref{st})] of nonlinear modes (\ref{solu2}).  Three-hump nonlinear modes (\ref{solu2}) for (b) $\sigma=0.2113$ (unbroken linear $\PT$-symmetry) and (d) $\sigma=0.22$ (broken linear $\PT$-symmetry).  (c) Stable and (e) unstable propagation of the nonlinear modes  (\ref{solu2}) corresponding to the weakly perturbed initial conditions shown in (b) and (d), respectively. Other parameters are $g=\omega=1,\, \Gamma=0.0375, \,  n=2$.}
 	\label{stability-W2-0.0375}
 \end{figure}

 \begin{figure}[!t]
 	\begin{center}
 	\vspace{0.05in}
 	\hspace{-0.05in}{\scalebox{0.4}[0.35]{\includegraphics{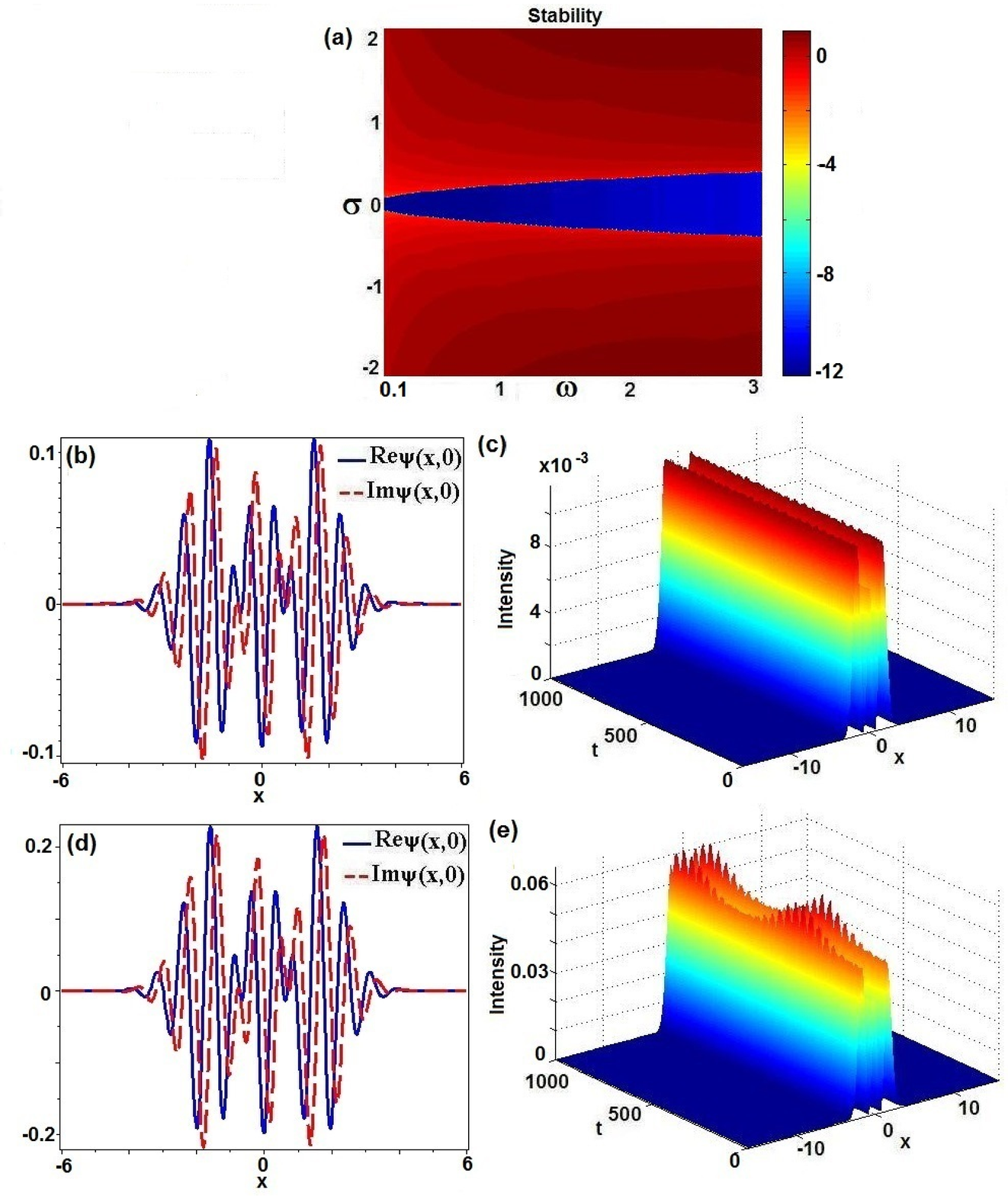}}}
 	\end{center}
 	\vspace{-0.15in} \caption{\small (color online). (a) Linear stability [cf. Eq.~(\ref{st})] of nonlinear modes (\ref{solu2}).
three-hump nonlinear modes (\ref{solu2}) for (b) $\sigma=0.1$ (unbroken linear $\PT$-symmetry) and (d) $\sigma=0.2113$ (unbroken linear $\PT$-symmetry). (c) Stable and (e) unstable propagation of the nonlinear modes (\ref{solu2}) corresponding to the weakly perturbed initial conditions shown in (b) and (d), respectively. Other parameters are $g=\omega=1,\, \Gamma=8, \, n=2$.}
 	\label{stability-W2-8}
 \end{figure}

For the case $n=2$, we firstly fix the momentum coefficient $\Gamma=0.0375$ and give the domain of linear stability [cf. Eq.~\ref{st})] of nonlinear mode (\ref{solu2}) in the $(\omega, \sigma)$-space (see  Fig.~\ref{stability-W2-0.0375}(a)). Moreover, we present the wave propagations of nonlinear modes is exhibited in Fig.~\ref{stability-W2-0.0375} via the direct propagation of the initially stationary state in
Eq.~(\ref{solu2}) with a noise perturbation of order $1\%$. In Fig.~\ref{stability-W2-0.0375}(c) we show that the nonlinear mode is stable where the linear $\PT$-symmetric phase is unbroken. If we increase the amplitude $\sigma$ of the gain-and-loss distribution a little bit (e.g., $\sigma=0.22)$ such that the nonlinear mode becomes unstable (see Fig.~\ref{stability-W2-0.0375}(d)).
We now change the momentum coefficient as $\Gamma=8$ and give the domain of linear stability [cf. Eq.~\ref{st})] of nonlinear mode (\ref{solu2}) in the $(\omega, \sigma)$-space (see  Fig.~\ref{stability-W2-8}(a)). Moreover, we present the wave propagations
 of nonlinear modes is exhibited in Fig.~\ref{stability-W2-8} via the direct propagation of the initially stationary state in Eq.~(\ref{solu2}) with a noise perturbation of order $1\%$. In Fig.~\ref{stability-W2-8}(c) we show that
the nonlinear mode is stable where the linear $\PT$-symmetric phase is unbroken. If we increase the amplitude $\sigma$ of the gain-and-loss distribution (e.g., $\sigma=0.2113$) even if the linear $\PT$-symmetric phase is still unbroken, but we have the unstable nonlinear mode (see Fig.~\ref{stability-W2-8}(e)). Particularly, it follows from Figs.~\ref{stability-W2-0.0375}(c) and \ref{stability-W2-8}(e) that for the fixed parameters $n=2,\, \omega=g=1,\, \sigma=0.2113$, the nonlinear mode (\ref{solu2}) is stable for $\Gamma=0.0375$, but if we increase $\Gamma$, then the nonlinear mode (\ref{solu2}) possibly becomes excited to the unstable state (e.g, $\Gamma=8$, see Fig.~\ref{stability-W2-8}(e)).

\vspace{0.2in} \centerline{\it 2.3 \, Non-$\PT$-symmetric nonlinear modes $(n=1)$}
\vspace{0.1in}

\begin{figure}[!t]
 	\begin{center}
 	\vspace{0.05in}
 	\hspace{-0.05in}{\scalebox{0.4}[0.4]{\includegraphics{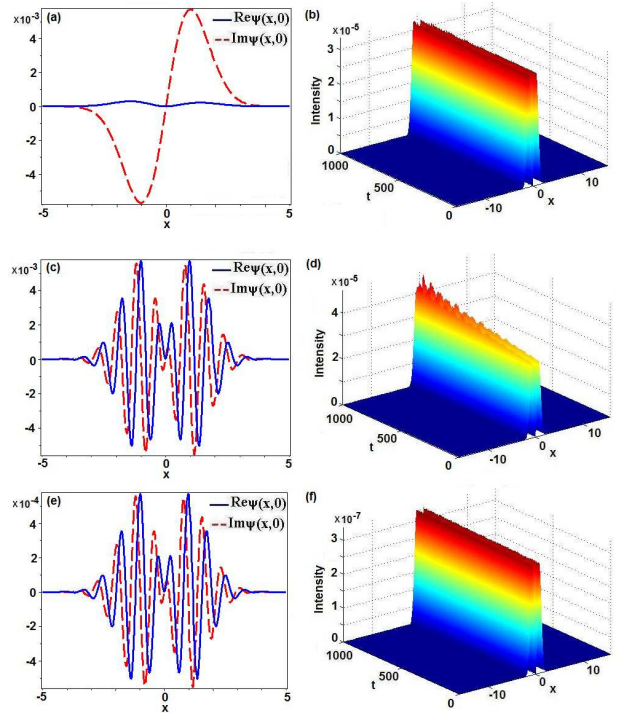}}}
 	\end{center}
 	\vspace{-0.15in} \caption{\small (color online). Double-hump (a) [$\sigma=0.01,\, \Gamma=0.0375$], (c) [$\sigma=0.01,\, \Gamma=8$], and  (e) [$\sigma=0.001,\, \Gamma=8$] nonlinear modes (\ref{solu2}). (b, f) Stable and (d) unstable propagation of the nonlinear modes described
by Eq.~(\ref{solu2}) corresponding to the weakly perturbed initial conditions shown in (a, e) and (c), respectively. Other parameters are
$g=\omega=n=1$.}
 	\label{stability-W1}
 \end{figure}

For the case $n=1$, where the complex potential is non-$\PT$-symmetric,  we have the linear stability analysis of nonlinear modes exhibited in Fig.~\ref{stability-W1} via the direct propagation of the initially stationary state in Eq.~(\ref{solu2}) with a noise perturbation of order $1\%$. In Fig.~\ref{stability-W1}(b) we show that the nonlinear mode is stable. If we fix the amplitude $\sigma=0.01$ of the gain-and-loss distribution and increase the momentum coefficient $\Gamma=8$ such that we have the unstable nonlinear mode (see Fig.~\ref{stability-W1}(d)). If we fix the momentum coefficient $\Gamma=8$ and decrease the amplitude $\sigma=0.001$  such that we have the stable nonlinear mode again (see Fig.~\ref{stability-W1}f).

 \begin{figure}[!t]
 	\begin{center}
 	\vspace{0.05in}
 	\hspace{-0.05in}{\scalebox{0.4}[0.4]{\includegraphics{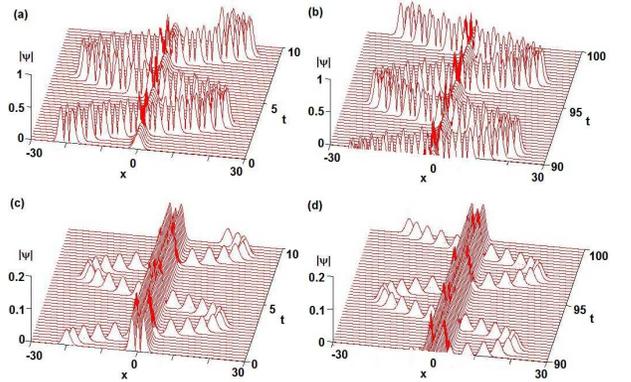}}}
 	\end{center}
 	\vspace{-0.15in} \caption{\small (color online). (a, b) the interaction of exact one-hump solution (\ref{solu2}) with $n=0$ and the wave $\sqrt{2}\sigma/3[4\omega(x+20)^2-2]\exp[-\omega(x+20)^2/2+4ix]$ with
 $\sigma=0.5$, (c, d) the interaction of exact three-hump solution (\ref{solu2}) with $n=2$ and the wave $\sqrt{2}\sigma/3\exp[-\omega(x+20)^2/2+4ix]$  with $\sigma=0.1$. Other parameters are $g=\omega=1, \, \Gamma=0.0375$. }
 	\label{osolu2}
 \end{figure}

 \vspace{0.2in} \centerline{\it 2.4 \, Interactions of solitons}
\vspace{0.1in}

 Moreover, we also study the interactions of the bright solitons (\ref{solu2}) with $n=0$ or $2$ with other nonlinear waves in the $\PT$-symmetric potential. For the case $n=0$, i.e., the solution is one-hump soliton,  we consider the initial condition $\psi(x,0)=\phi_0(x)+\sqrt{2}\sigma/3[4\omega(x+20)^2-2]\exp[-\omega(x+20)^2/2+4ix]$ with $\phi_0(x)$ given by Eq.~(\ref{solu2}) such that the elastic interaction is generated (see Figs.~\ref{osolu2}(a, b)).
 For the case $n=2$, i.e., the solution is three-hump soliton,  we consider the initial condition $\psi(x,0)=\phi_2(x)+\sqrt{2}\sigma/3\exp[-\omega(x+20)^2/2+4ix]$ with $\phi_2(x)$ given by Eq.~(\ref{solu2}) such that the elastic interaction is generated (see Figs.~\ref{osolu2}(c, d)).

\subsubsection{The  varying transverse power-flow}

To more understand the properties of stationary nonlinear mode (\ref{solu2}), we check its transverse power-flow or `Poynting vector' as
\bee \label{sn}
S_n(x)\!=\!\frac{2\sigma^2}{9g}\!H_n^2(\sqrt{\omega}x)e^{-\omega x^2}\!\!\left[\Gamma\!-\!\frac{2\sigma}{3}H_n(\sqrt{\omega}x)e^{-\omega x^2/2}\right].\,\,\,\,\,\,
\ene
which denotes that the signs of $S_n$ are not positive or negative definite and dependent on parameters $\sigma$ and $\Gamma, \, n$, and even space position $x$. For $n=0$, the three results are listed in Table~\ref{table-x}. We discuss them as follows:

  i) For the case 1, the power always flows from the loss toward the gain.

ii)  For the case 2, the the power flowing directions  are complicated. For any $0<\Gamma<2\sigma/3$ with $\sigma>0$,
 the power always flows from the loss toward the gain for $|x|\geq \sqrt{-\frac{2}{\omega}\ln\left(\frac{3\Gamma}{2\sigma}\right)}$ and from the gain toward the loss for $|x| < \sqrt{-\frac{2}{\omega}\ln\left(\frac{3\Gamma}{2\sigma}\right)}$.

  iii)  For the case 3, we find that the power always flows from the gain toward the loss, which displays that the rotational operator $i\Gamma\partial_x$ is the same (or weak) action as the gain-and-loss term $W(x)$ ($\Gamma\not=0$) or has no effect on gain-and-loss term (i.e., $\Gamma=0$).

\begin{table}[!t]
\vspace{-0.1in}
\caption{\small The signs of $S_0(x)$ (the power flowing direction) is related to $\sigma$, $\Gamma$, and even space. \vspace{0.05in}}
\begin{tabular}{ccll} \hline\hline \\ [-2.0ex]
 Case &    $\sigma$   & \qquad $\Gamma$   &  \quad $S_0(x)$ \\ [1.0ex] \hline \\ [-2.0ex]
1 &  $\sigma>0$  \,\,  & $\Gamma\geq \frac{2\sigma}{3}$  & $S_0\geq 0$ \,\, {\rm for~all~} $x$ \\ [1.0ex] \hline \\ [-2.0ex]
2&  $\sigma>0$ \,\, & $0<\Gamma<\frac{2\sigma}{3} $ \qquad &
               $\begin{cases}S_0\geq 0, \,\, \,\, |x|\!\geq\! \sqrt{-\frac{2}{\omega}\ln\left(\frac{3\Gamma}{2\sigma}\right)} \vspace{0.05in}\\
                               S_0< 0, \,\, \,\ otherwise
                                                               \end{cases}$     \\ [1.0ex] \hline \\ [-2.0ex]
3&  $\sigma<0$ \,\,  & $\Gamma >0$  & $S_0> 0$ \,\, {\rm for~all~} $x$    \\ [1.0ex] \hline\hline
\end{tabular}
\label{table-x}
\end{table}

\begin{figure}[!t]
	\begin{center}
	\vspace{0.1in}{\scalebox{0.22}[0.15]{\includegraphics{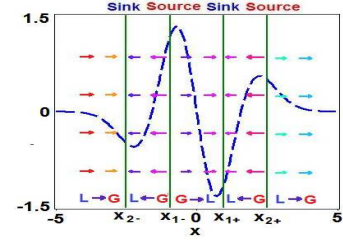}}}
				\end{center}
	\vspace{-0.2in}
	\caption{ (color online). The transverse power-flow or `Poynting vector' $S_2(x)$ (\ref{sn}) related to nonlinear modes (\ref{solu2}) and gain-and-loss curve $W_2(x)$ given by Eq.~(\ref{psi2}) (dashed lines) with $n=2,\, \sigma=0.3,\, \Gamma=0.2,\, \omega=1.$\, $x_{1\pm}\approx\pm 0.9434788009,\, x_{2\pm}\approx\pm  2.506418670$. `L'=`Loss', `G'=`Gain'.}	 \label{fig-vector-ns}
\end{figure}

For $n=1$, the transverse power-flow or `Poynting vector' $S_1(x)$ admits the similar structure as $S_0(x)$. But as $n>1$, $S_n(x)$
admits the complicated structures. For example, when we choose $n=2$, $\Gamma=0.2,\, \sigma=0.3$, it is difficult to find exact roots of equation $f_2(x)=0.2-0.4(2x^2-1)e^{-x^2/2}=0$, but we have its four numerical roots $x_{1\pm}\approx\pm 0.9434788009,\, x_{2\pm}\approx\pm  2.506418670$. As a result, we know that $S_2(x)\geq 0$ for $|x|\geq 2.506418670$ or $|x|\leq 0.9434788009$ and $S_2(x)\leq 0$ for $0.9434788009\leq |x| \leq 2.506418670$ such that we have the graph to exhibit the relation between transverse power-flow and gain-and-loss distribution (see Fig.~\ref{fig-vector-ns}).

\begin{figure}[!t]
 	\begin{center}
 	\vspace{0.05in}
 	\hspace{-0.05in}{\scalebox{0.4}[0.4]{\includegraphics{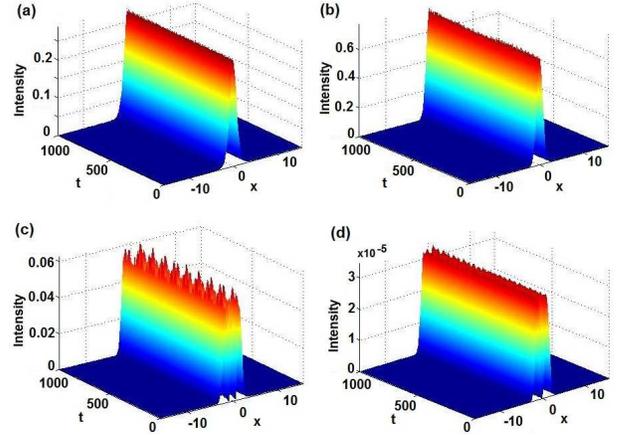}}}
 	\end{center}
 	\vspace{-0.15in} \caption{\small (color online). The propagation dynamics with a higher noise level of $3\%$. (a) the parameters are as ones in Fig.~\ref{stability-g1}d, (b) the parameters are as ones in Fig.~\ref{stability-W0-0.0375}c, (c) the parameters are as ones in Fig.~\ref{stability-W2-0.0375}c, (d) the parameters are as ones in Fig.~\ref{stability-W1}b. }
  \label{hn}
 \end{figure}

In the above-mentioned propagation dynamics of nonlinear modes, we use the a small noise level of $1\%$. We here choose
some examples (e.g., Figs.~\ref{stability-g1}d, \ref{stability-W0-0.0375}c, \ref{stability-W1}c, and  \ref{stability-W2-0.0375}b) to repeat the evolutions with a higher noise level of $3\%$ (see Fig.~\ref{hn}) such that we find that a higher noise level may lead to the unobvious (see Figs.~\ref{hn}a and b) and obvious
(see Figs.~\ref{hn}c and d) periodic oscillations rather than `destroy' their  stability.

\section{The equation (\ref{nls}) with varying momentum coefficient }

Now we turn to Eq.~(\ref{nls}) with the varying momentum coefficient as
\bee\label{ga}
\Gamma(x)=\gamma\, {\rm sech} x
\ene
with $\gamma$ being a real parameter, and the $\PT$-symmetric potential $V(x)+iW(x)$ is still chosen as the Scarff-II potential (\ref{ps}).

\subsection{Linear problem with unbroken and broken $\PT$-symmetry}

The linear eigenvalue problem with $\PT$-symmetric Scarff-II potential (\ref{ps}) related to Eq.~(\ref{nls}) is written as
\bee \label{lsv}
 L_{v}\Phi\!=\!\lambda\Phi,\, L_{v}\!=\!-\frac{1}{2}\partial_x^2\!+\!i\Gamma(x)\partial_x\!+\! V(x)\!+\! iW(x),\,\,\,\,\,
\ene
where $\lambda$ and $\Phi(x)$ are eigenvalue and engenfunction, respectively, and $V(x),\, W(x),\, \Gamma(x)$ are defined by Eqs.~(\ref{ps}) and (\ref{ga}).  In the absence of the momentum term $\gamma=0$, we know that $L_v$ reduces to the usual $\PT$-symmetric Hamiltonian operator $L_0$ and admits entirely real spectra provided that the parameters $V_0<0$ and $W_0$ satisfy the condition (\ref{ptc})~\cite{scarff}.

In the presence of momentum term $\gamma$, we make the invertible transformation in Eq.~(\ref{lsv})
\bee\label{tr}
\Phi(x)=\hat\Phi(x)\exp[i\gamma\tan^{-1}(\sinh x)].
 \ene
 such that we find that $\hat\Phi(x)$ is defined by
\bee \label{lsvg}
 L_{vc}\hat\Phi(x)=\lambda\hat\Phi(x),\quad
  L_{vc}\!=\!-\frac{1}{2}\partial_x^2+U(x),
\ene
where we have introduced the complex potential $U(x)$ as
\bee\begin{array}{rl}
U(x)\!&\!=\!
\left(V_0\!-\!\dfrac{\gamma^2}{2}\right){\rm sech}^2 x\!+\!i\left(W_0\!+\!\dfrac{\gamma}{2}\right){\rm sech}x\tanh x,
\end{array}\quad
\ene
which is just a $\PT$-symmetric Scarff-II potential. It follows from Eqs.~(\ref{lsv}) and (\ref{lsvg}) that two linear operators $L_{v}$ and $L_{vc}$ admit the same spectra via the transformation (\ref{tr}).  Thus we know that the linear problems (\ref{lsv}) and (\ref{lsvg}) admits entirely real (discrete) spectra provided that the parameters $V_0<\gamma^2/2$ and $W_0$ satisfy
\bee
\left|W_0+\frac{\gamma}{2}\right|\leq \frac{\gamma^2}{2}+\frac{1}{8}-V_0,
\ene
that is, two families of lines for every  $\gamma$
\bee
l_{\gamma,\pm}: W_0=\pm \left(\frac{\gamma^2}{2}+\frac{1}{8}-V_0\right)-\frac{\gamma}{2}
\ene
are $\PT$-symmetric threshold lines in $(V_0, W_0)$-space.

 For any real $\gamma$, we find that all lines $l_{\gamma, +}$ in upper half plane are parallel except for
 $l_{0,+}=l_{1,+}$ and all lines $l_{\gamma, -}$ in lower half plane are also parallel except for $l_{0,-}=l_{-1,-}$,
 the critical line $l_{0.5, +}$ is lowest in all lines $l_{\gamma, +}$ and $l_{-0.5, -}$ is highest in all lines $l_{\gamma, -}$. Since we know that the distance between two crossover points of $l_{\gamma, \pm}$ and $V_0=0$ is $d_{\gamma}=\gamma^2+0.25\geq 0.25$, then the domains of unbroken $\PT$-symmetric phase become bigger and bigger as $|\gamma|$ increase, that is, the domain of unbroken $\PT$-symmetric phase is smallest for $\gamma=0$. This can be understood from the fact that the decreasing of $\gamma {\rm sech} x$ corresponds to the shrinking of the gain-and-loss domains.

\begin{figure}[!t]
 	\begin{center}
 	\vspace{0.05in}
 	\hspace{-0.05in}{\scalebox{0.3}[0.3]{\includegraphics{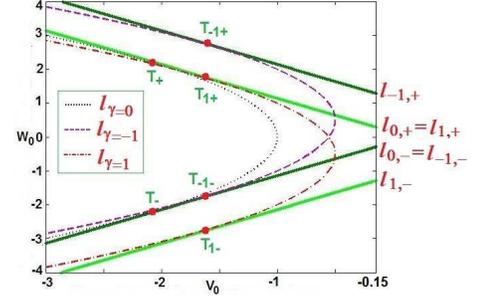}}}
 	\end{center}
 	\vspace{-0.15in} \caption{\small (color online). The profile of phase transitions for the linear operator $L_v$ (\ref{lsv}) with varying frequency (\ref{ga}) and $\PT$-symmetric Scarff-II potential (\ref{ps}). The unbroken (broken) $\PT$-symmetric phase is in the domain between (outside) two symmetric phase breaking lines for every different frequency $\gamma=0, \pm 1$. The existence conditions of bright soliton (\ref{solux}) for the attractive $g=1$ (repulsive $g=-1$) cases is in the domain outside (inside) the parabolic curve $V_0=-[2(W_0^2+\gamma W_0-2\gamma^2)/9+1]$ for $\gamma=0, \pm 1$. The tangent points are $T_{\pm}=(-2.125,\, \pm 2.25),\, T_{1\pm}=(-1.625,\, \pm 2.25-0.5),\, T_{-1\pm}=(-1.625,\, \pm 2.25+0.5)$. }
 	\label{fig-spec-sech}
 \end{figure}

\begin{figure}[!ht]
	\begin{center}
		\hspace{-0.05in}{\scalebox{0.35}[0.35]{\includegraphics{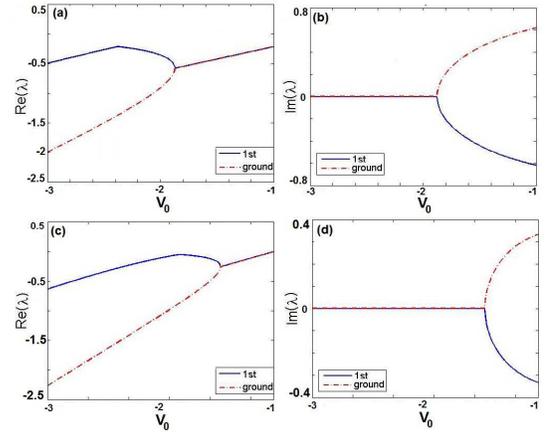}}}
				\end{center}
	\vspace{-0.15in}
	\caption{ (Color online).  (a, c) Real and (b, d) imaginary parts of the eigenvalues $¦Ë$ [see Eq.~(\ref{lsv})]
as functions of $V_0<0$ for the potential (\ref{ps}) at (a, b) $W_0=-3$ and (c, d) $W_0=1$, as well as $\gamma=1$.}		
\label{fig1i-sech}
\end{figure}

We also numerically give the domains of the unbroken and broken $\PT$-symmetry phase of the rotating operator $L_{v}$ with $\PT$-symmetric potential (\ref{ps}) and varying frequency (\ref{ga}) for the different amplitude $\gamma$ on the $(W_0, V_0)$-space (see Fig.~\ref{fig-spec-sech}).   For the given $\gamma=1$ and different amplitude $W_0=-3, 1$, we numerically illustrate two lowest states corresponding to discrete spectra such that the spontaneous symmetry breaking occurs due to collision of the two lowest states as the amplitude $|V_0|$ of the potential decreases (see Fig.~\ref{fig1i-sech}). The non-smooth point is due to the order of energy levels (i.e, the real parts of eigenvalues) (see Fig.~\ref{fig1i-sech}a).

\subsection{Nonlinear modes and stability}

We now study nonlinear stationary modes of Eq.~(\ref{nls}) in $\PT$-symmetric Scarff-II potential (\ref{ps}) with $V_0<0$ and $W_0\in \mathbb{R}$, whose bright solitons can be found by $\psi(x,t)=\phi(x)e^{i\mu t}$ with
\bee\label{solux}
 \phi(x)\!=\!\sqrt{\frac{1}{g}\!\left[\frac{2}{9}(W_0^2\!+\!\gamma W_0\!-\!2\gamma^2)\!+\!V_0\!+\!1\right]}\,{\rm sech}x\,e^{i\varphi(x)} \nonumber \\
 \hspace{-0.2in}\ene
for both cases $g=\pm 1$, where the chemical potential is $\mu=0.5$, and the phase is related to the amplitude $W_0$ of gain-and-loss distribution and frequency $\gamma$, that is
$\varphi(x)=\frac{2(\gamma-W_0)}{3}\tan^{-1}(\sinh x).$
The existence conditions of the bright solitons (\ref{solux}) are
$V_0>-[2(W_0^2+\gamma W_0-2\gamma^2)/9+1]$ for $g=1$ and $V_0<-[2(W_0^2+\gamma W_0-2\gamma^2)/9+1]$
   for $g=-1$. Thus the family of parabolas
\bee
 l_{\gamma}:\, V_0=-[2(W_0^2+\gamma W_0-2\gamma^2)/9+1] \ene
 with the  vertex being  $(V_0, W_0)=(0.5\gamma^2-1,\, -0.5\gamma)$ can be regarded as a critical curve for existence conditions of bright solitons (\ref{solux})  for $g=\pm 1$. Moreover, for any $\gamma$, we find an interesting result that the parabola $l_{\gamma}$ is tangent to two critical lines $l_{\gamma, \pm}$ for the $\PT$-symmetric phase shift with the tangent points are $(-2.125+0.5\gamma^2,\, \pm 2.25-0.5\gamma)$ for any $\gamma$. That is, the domain of bright solitons (\ref{solux}) for $g=-1$ is located inside the domain of unbroken $\PT$-symmetry. The domain of bright solitons (\ref{solux}) for $g=1$ contains partial domain of unbroken $\PT$-symmetry and all domain of broken $\PT$-symmetry. Particularly, the two parabolas $l_{\gamma=1}$ and $l_{\gamma=0}$ are both tangent to the same line $l_{1, +}=l_{0, +}$ with different tangent points, and two parabolas $l_{\gamma=-1}$ and $l_{\gamma=0}$ are both tangent to the same line $l_{-1, -}=l_{0, -}$  with different tangent points (see Fig.~\ref{fig-spec-sech}).

\begin{figure}[!t]
 	\begin{center}
 	\vspace{0.05in}
 	\hspace{-0.05in}{\scalebox{0.4}[0.35]{\includegraphics{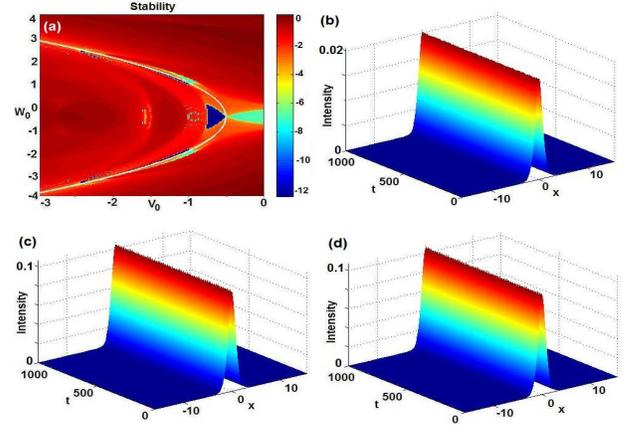}}}
 	\end{center}
 	\vspace{-0.15in} \caption{\small (color online)  Linear stability [cf. Eq.~(\ref{st})] of nonlinear modes (\ref{solux}) for (a) $g=1$ and $\gamma=1$.  Stable  propagation of the nonlinear modes (\ref{solux}) for (b) $V_0=-0.5,\, W_0=-0.8$ (unbroken linear $\PT$-symmetry), (c) $V_0=-1,\, W_0=1.15$ (broken linear $\PT$-symmetry), and (d) $V_0=-1,\, W_0=-2.15$ (broken linear $\PT$-symmetry).} \label{stability-g1-sech}
 \end{figure}

\begin{figure}[!t]
 	\begin{center}
 	\vspace{0.05in}
 	\hspace{-0.05in}{\scalebox{0.4}[0.4]{\includegraphics{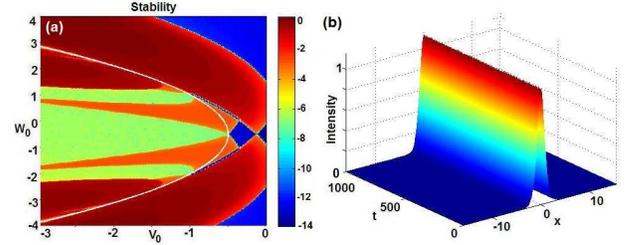}}}
 	\end{center}
 	\vspace{-0.15in} \caption{\small (color online)   Linear stability [cf. Eq.~(\ref{st})]  of nonlinear modes (\ref{solux}) for (a) $g=-1$ and $\gamma=1$. Stable propagation of the nonlinear modes
  (\ref{solux}) for (b) $V_0=-2,\, W_0=-1.8$ (unbroken linear $\PT$-symmetry).} \label{stability-g-1-sech}
 \end{figure}

For the fixed $\gamma=1$, Fig.~\ref{stability-g1-sech}a for $g=1$ and Fig.~\ref{stability-g-1-sech}a for $g=-1$ display the maximal absolute values of imaginary parts of the linearized eigenvalues $\delta$ related to solutions (\ref{solux}) as a function of $V_0$ and $W_0$ [cf. Eq.~(\ref{st})], which illustrate the linear stable (the dark blue) and other unstable regions.

In the following, we numerically check the robustness of nonlinear modes (\ref{solux}) for both attractive and repulsive cases via the direct propagation of the initially stationary state in Eq.~(\ref{solux}) with a noise perturbation of order $1\%$. For the attractive case $g=1$ and $\gamma=1$, Fig.~\ref{stability-g1-sech} illustrates the stable and unstable situations for different parameters $W_0$ and $V_0$. For $V_0=-0.5$ and $W_0=-0.8$ belonging to the domain of the unbroken linear $\PT$-symmetric
phase of the operator $L_v$ [cf. Eq.~(\ref{lsv})], the nonlinear mode is stable (see Fig.~\ref{stability-g1-sech}(b)). If we fix $V_0=-1$ and change $W_0=1.15,\, -2.15$, which even if corresponds to the domain of the broken linear $\PT$-symmetric phase of the operator $L_v$, the nonlinear mode still becomes stable (see Figs.~\ref{stability-g1-sech}(c, d)).
For the repulsive case $g=-1$ and $\gamma=1$, Fig.~\ref{stability-g-1-sech}(b) illustrates the stable mode for $V_0=-2$ and $W_0=-1.8$ related to the unbroken linear $\PT$-symmetric phase of $L_v$.

Finally, we also illustrate the transverse power-flow or `Poynting vector' of the nonlinear mode (\ref{solux}) as
\bee
S(x)=\frac{2}{3}(\gamma-W_0)\phi_0^2\, {\rm sech}^3 x,
\ene
where $\phi_0=\sqrt{\frac{1}{g}\left[V_0+1+\frac{2}{9}(W_0^2+\gamma W_0-2\gamma^2)\right]}$.
When $\gamma>0>W_0$ or $W_0>\gamma>0$, the power flows from the gain toward loss, whereas $\gamma>W_0>0$ the power flows from the loss toward gain.

\section{3D model with $\PT$-symmetric potential}

We consider the 3D generalized GP equation with the $\PT$-symmetric potential
\bee\label{nls3}
 i\psi_t\!=\!\left(\!-\frac12\nabla^2+i\Gamma\cdot\partial_{\br}+V(\br)+iW(\br)-g|\psi|^2\!\right)\!\psi,\,\,\,\,\,
\ene
where $\br=(x,y,z)$, $\nabla^2=\partial_x^2+\partial_y^2+\partial_z^2$, ${\bf \Gamma}=(\Gamma_x, \Gamma_y, \Gamma_z)$ with $\Gamma_{x,y,z}$ being real constants, and $\partial_{\br}=(\partial_x, \partial_y,\partial_z)$.

We focus on the stationary solution of Eq.~(\ref{nls3}) $\psi(\br,t)=\phi(\br)e^{i\mu t}$ and the field $\phi(\br)$ satisfies the stationary model
\bee\label{nls3o}
 \left(\!-\!\frac12\nabla^2+i{\bf \Gamma\cdot\partial}+V(\br)+iW(\br)-g|\phi|^2+\mu\!\right)\!\phi=0,\qquad
\ene

 \begin{figure}[!t]
 	\begin{center}
 	\vspace{0.05in}
 	\hspace{-0.05in}{\scalebox{0.42}[0.35]{\includegraphics{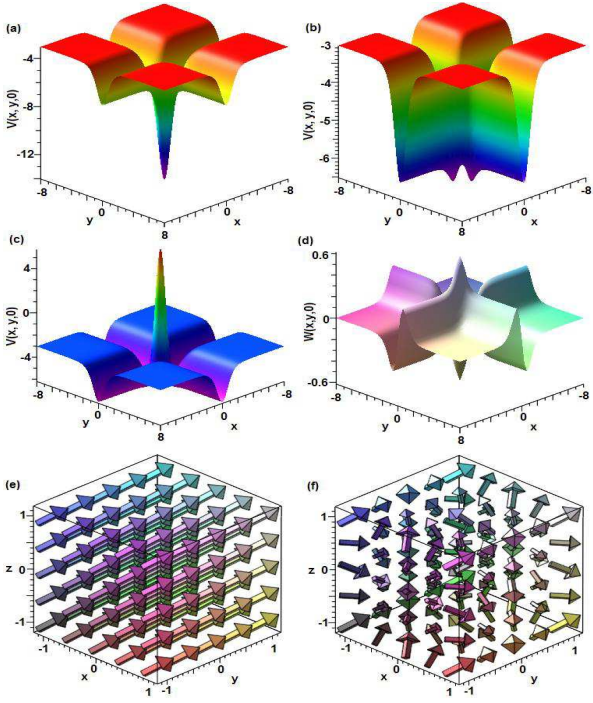}}}
 	\end{center}
 	\vspace{-0.15in} \caption{\small (color online). The potential $V(x,y,0)$ given by Eq.~(\ref{potend})  with (a) $v_0=-5$, (b) $v_0=5$, and (c) $v_0=15$; (d) the gain-and-loss
distribution $W(x,y,0)$ given by Eq.~(\ref{potend}).  The velocity filed with (e) $\Gamma_{\eta}=3$ and (f) $\Gamma_{\eta}=0.2$. Other parameters are $\alpha_{\eta}=2$ and $W_0=0.8$.}
 	\label{3d-ri}
 \end{figure}

 \begin{figure}[!t]
 	\begin{center}
 	\vspace{0.05in}
 	\hspace{-0.05in}{\scalebox{0.42}[0.35]{\includegraphics{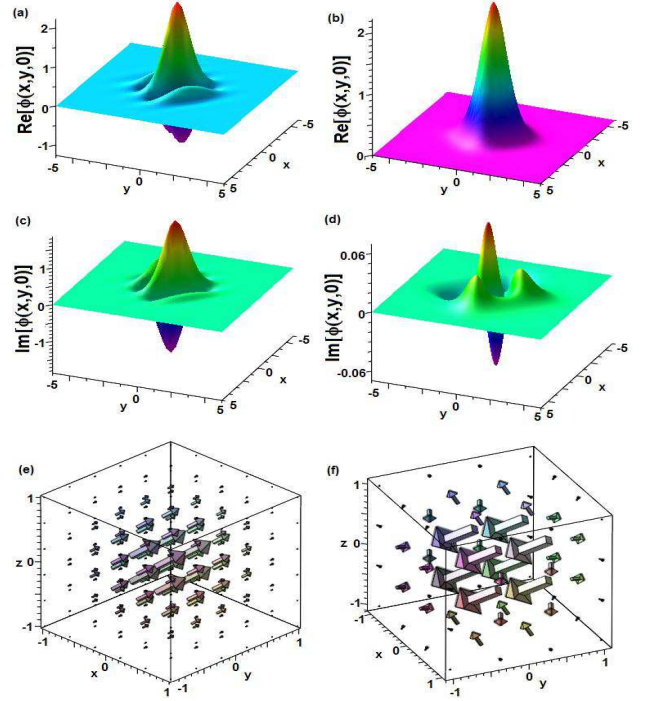}}}
 	\end{center}
 	\vspace{-0.15in} \caption{\small (color online). The (a, b) real parts and (c,d) imaginary parts
    of the bright soliton solution (\ref{solu4}); (e, f) the transverse power flow vector (Poynting vector) (\ref{3d-vector}).
    Parameters are $\Gamma_{\eta}=3$ (left column) and $\Gamma_{\eta}=0.2$ (right column). Other parameters are $g=1,\, v_0=5,\, W_0=0.8$,  and $\alpha_{\eta}=2$. }
 	\label{3d-pri}
 \end{figure}

Nowadays we consider the generalized 3D $\PT$-symmetric Scarff-II potential
\bee\label{potend} \begin{array}{l}
 V(\br)\!=\!\sum_{\eta}\!\left(v_{1\eta}{\rm sech}^2\eta\!+\!v_{2\eta} {\rm sech}^{2\alpha_{\eta}}\eta\!\right)
  \!+\!v_0\!\prod_{\eta}{\rm sech}^{2\alpha_{\eta}}\eta, \vspace{0.1in}\\
 W(\br)=W_0\sum_{\eta}{\rm sech}^{\alpha_{\eta}}\eta\tanh\eta,
\end{array}\ene
where the index $\eta=x,y,z$,\, $v_{1\eta}=-\alpha_{\eta}(\alpha_{\eta}+1)/2$, $v_{2\eta}=-2W_0^2/(9\alpha_{\eta}^2)$, $\alpha_{\eta}>0$,  $v_0\not=0,\, W_0$  are all real constants. Figs.~\ref{3d-ri}(a)-(d) illustrate the profiles of $\PT$-symmetric potentials $V(x,y,0)$ and $W(x,y,0)$.

For the above-mentioned 3D $\PT$-symmetric potential (\ref{potend}) we obtain exact bright solitons of Eq.~(\ref{nls3o})
\bee\label{solu4}
\phi(\br)\!=\!\sqrt{v_0/g}\prod_{\eta}{\rm sech}^{\alpha_{\eta}}\eta\, e^{i\varphi(\br)},
\ene
where $gv_0>0$, the chemical potential is $\mu=\sum_{\eta}(\alpha_{\eta}^2+\Gamma_{\eta}^2)/2$ and the phase is
 \bee \nonumber
\varphi(\br)=\sum_{\eta}\left({\bf\Gamma}\cdot\br-\frac{2W_0}{3\alpha_{\eta}}\int^{\eta}_0{\rm sech}^{\alpha_{\eta}}sds\right).
 \ene
The real and imaginary parts of the solutions (\ref{solu4}) are shown in Figs.~\ref{3d-pri}(a)-(d).

The velocity field $\bv(x,y,z)$ of the solitons (\ref{solu4}) have the form
$\bv=\nabla\varphi(x,y,z)=(f_x, f_y, f_z)$, where $f_{\eta}=\Gamma_{\eta}-2W_0/(3\alpha_{\eta}){\rm sech}^{\alpha_{\eta}}{\eta}$,
which is shown in Figs.~\ref{3d-ri}(e) and (f). Therefore, the divergence of velocity field $\bv(x,y,z)$ (alias the flux density) is given by
\bee \label{div}
\begin{array}{l}
{\rm div}\,\bv(x,y,z)=\nabla^2\varphi(x,y,z)=\frac{2}{3}W(\br), \end{array}
\ene
which measures the flux per unit area and is dependent on both $W_0$ and space position. Moreover, we also have the divergence of velocity field is proportional to the gain-and-loss distribution $W(\br)$ (cf. Fig.~\ref{3d-ri}(d)).

From Eq.~(\ref{div}) we have the following proposition:
\begin{itemize}
\item{} the direction of a flux is away from the point if $W(\br)>0$ for a point, i.e., $W(\br)$ is the gain term;

\item{}  the direction of a flux is towards the point if $W(\br)<0$ for a point, i.e., $W(\br)$ is the loss term;

\item{}  a flux does not move at the point if $W(\br)=0$ for a point, i.e, without the gain-and-loss term,
\end{itemize}

The transverse power-flow or Poynting vector related to the solution (\ref{solu4}) is given by
\bee \label{3d-vector}
 \vec{S}(\br)=
 \frac{v_0}{g}\left(\prod_{\eta}{\rm sech}^{2\alpha_{\eta}}\eta\right)\!(f_x, f_y, f_z),\,\,\ene
 which illustrates the complicated structures whose profiles are displayed in Figs.~\ref{3d-pri}(e) and (f) for some chosen parameters.

In fact, we can also consider Eq.~(\ref{nls3}) with varying coefficient $\Gamma\to \Gamma(\br)$, which will be studied in another literature.

\section{conclusions}

In conclusion, we have presented bright solitons and their linear stability of the generalized GP equation with physically relevant $\PT$-symmetric potentials. We find that the constant momentum coefficient has no effect on the unbroken linear $\PT$-symmetric phase, but it can enlarge real parts of the spectra, and modulate stability and the transverse power-flow of nonlinear modes. If we consider the spatially varying  momentum coefficient, we find that $\Gamma(x)$ not only changes the unbroken linear $\PT$-symmetric phase but controls stability of nonlinear modes. Moreover, the elastic interactions of two bright solitons are illustrated in the $\PT$-symmetric potentials.

Particularly, the nonlinearity can modulate the linear modes with broken $\PT$-symmetric phases into stable nonlinear modes.
 We also consider the stability of bright solitons in the presence of non-$\PT$-symmetric potential (i.e., harmonic-Gaussian potential given by Eqs.~(\ref{ps2a}) and (\ref{ps2b}) with $n=1$).  Finally, we study nonlinear modes in 3D generalized GP equation with the generalized $\PT$-symmetric Scarff-II potential. These stable PT-symmetric nonlinear modes provide the
abundant data to design the relevant physical experiments and may excite the applications in some related fields. The methods used in this paper can also be extended to other nonlinear models with $\PT$-symmetric or non-$\PT$-symmetric potentials.

\acknowledgments

The authors would like to thank the referees for their valuable suggestions that have improved substantially the
paper. This work was supported by the NSFC under Grant No. 11571346 and the Youth Innovation Promotion Association CAS.


\begin{thebibliography}{99}

\bibitem{qm} G. Barton, {\em Introduction to Advanced Field Theory} (New York: Wiley, 1963).

\bibitem{Bender98} C. M. Bender and S. Boettcher, Phys. Rev. Lett. {\bf 80}, 5243 (1998).

\bibitem{Bender2} C. M. Bender, Rep. Prog. Phys. {\bf 70}, 947 (2007).


\bibitem{Exp1} A. Guo, {\it et al}., Phys. Rev. Lett. {\bf 103}, 093902 (2009).

\bibitem{Exp2} C. E. R\"uter, {\it et al}., Nature Phys. {\bf 6}, 192 (2010).

\bibitem{exp2a} Alois Regensburger, {\it et al.,} Nature {\bf 488}, 167 (2012).

\bibitem{Exp3} G. Castaldi, {\it et al.,}  Phys. Rev. Lett. {\bf 110}, 173901 (2013).

\bibitem{Exp4} A. Regensburger, {\it et al.,} Phys. Rev. Lett. {\bf 110}, 223902 (2013).

\bibitem{exp5} B. Peng, {\it et al.,} Nature Phys. {\bf 10}, 394 (2014).

\bibitem{ptsf} Z. H. Musslimani, {\it et al.,} Phys. Rev. Lett.  {\bf 100}, 030402 (2008); Z. H. Musslimani, {\it et al.,} J. Phys. A {\bf 41}, 244019 (2008).

\bibitem{yanpre15} Z. Yan,  {\it et al.}, Phys. Rev. E {\bf 92}, 022913 (2015).

\bibitem{ptsf2} F. K. Abdullaev,  {\it et al.}, Phys. Rev. A {\bf 83}, 041805(R) (2011); S. Nixon,  {\it et al.}, Phys. Rev. A {\bf 85}, 023822 (2012); N. Moiseyev, Phys. Rev. A, {\bf 83}, 052125 (2011);  Y. Lumer, {\it et al.,}  Phys. Rev. Lett. {\bf 111}, 263901 (2013).  C. P. Jisha, {\it et al.,}  Phys. Rev. A {\bf 89}, 013812 (2014).

\bibitem{harm1} Z.  Yan, {\it et al., } arXiv.1009.4023 (2010); Z. Yan,  Phil. Trans. R. Soc. A {\bf 371}, 20120059 (2013).

\bibitem{harm-l} D. A. Zezyulin and V. V. Konotop,  Phys. Rev. A {\bf 85}, 043840 (2012).

\bibitem{gau2} S. Hu, {\it et al., }  Phys. Rev.A {\bf 84}, 043818 (2011); V. Achilleos, {\it et al.}, Phys. Rev. A {\bf 86}, 013808 (2012); J. Yang, Opt. Lett. {\bf 39}, 5547 (2014).

\bibitem{anharm} Z. C. Wen and Z. an, Phys. Lett. A {\bf 379}, 2025 (2015).

\bibitem{yanpra15} Z. Yan,  {\it et al.}, Phys. Rev. A {\bf 92}, 023821 (2015).


\bibitem{delta} H. Cartarius and G. Wunner, Phys. Rev. A {\bf 86}, 013612 (2012); F. Single, Phys. Rev. A {\bf 90}, 042123 (2014).

\bibitem{sg} C. P. Jisha, {\it et al.,} Phys. Rev. A {\bf 90}, 043855 (2014).

\bibitem{other} C. Yin,  {\it et al.}, Opt. Exp. {\bf 20}, 19355 (2012); G. Burlak and B. A. Malomed, Phys. Rev. E {\bf 88}, 062904 (2013); Yu. V. Bludov,  {\it et al.}, Phys. Rev. A  {\bf 87}, 013816 (2013); R. Fortanier, {\it et al.,} Phys. Rev. A {\bf 89}, 063608 (2014);  D. Dizdarevic, {\it et al.,} Phys. Rev. A {\bf 91}, 033636 (2015).

\bibitem{yan16} Y. Chen and Z. Yan, Sci. Rep. {\bf 6}, 23478 (2016).

\bibitem{rbec} H. Saito and M. Ueda, Phys. Rev. Lett. {\bf 93}, 220402 (2004).

\bibitem{rbecp1} S. Schwartz, {\it et al.,} New. J. Phys. {\bf 8}, 162 (2006).

\bibitem{rbecp2} Y. Li,  {\it et al.}, Phys. Rev. A {\bf 86}, 023832 (2012).

\bibitem{rbecb} J. L. Helm,  {\it et al.}, Phys. Rev. Lett. {\bf 114}, 134101 (2015).

\bibitem{so1} Y. J. Lin, {\it et al.,} Nature {\bf 471}, 83 (2011).

\bibitem{so2} Y. Zhang, {\it et al.,} Phys. Rev. Lett. {\bf 108}, 035302 (2012).

\bibitem{so3} Y. V. Kartashov,{\it et al.,} Phys. Rev. A {\bf 90}, 063621 (2014).






\bibitem{stable} E. A. Kuznetsov,  {\it et al.},  Phys. Rep. {\bf 142}, 103 (1986).

\bibitem{yang} J. Yang, {\it Nonlinear Waves in Integrable and Nonintegrable Systems} (SIAM, Philadelphia, 2010).

\bibitem{scarff} M. Znojil, J. Phys. A {\bf 33}, L61 (2000); Z. Ahmed, Phys. Lett. A 282, 343 (2001).

\bibitem{rp} G. P\"oschl and E. Teller, Z. Phys. {\bf 83}, 143 (1993).







\end{thebibliography}
\end{document}